%% file: main.tex
\documentclass[sigconf,natbib=true,anonymous=false]{acmart}
\AtBeginDocument{%
  }

\setcopyright{acmlicensed}
\copyrightyear{2018}
\acmYear{2018}
\acmDOI{XXXXXXX.XXXXXXX}
\acmConference[Conference acronym 'XX]{Make sure to enter the correct
  conference title from your rights confirmation email}{June 03--05,
  2018}{Woodstock, NY}

\acmISBN{978-1-4503-XXXX-X/2018/06}
\usepackage{graphicx}
\usepackage{subfigure}
\usepackage{amsmath}
\usepackage{amsthm}
\usepackage{booktabs}
\usepackage{algorithm}
\usepackage{algorithmic}
\usepackage[switch]{lineno}
\usepackage{enumitem}
\usepackage{multirow}
\begin{document}

\title{Modeling Domain and Feedback Transitions for \\Cross-Domain Sequential Recommendation}

\author{Changshuo Zhang}
\authornote{Both authors contributed equally to this research.}
\affiliation{%
  \institution{Gaoling School of AI,\\ Renmin University of
China}
  \city{Beijing}
  \country{China}
}
\email{lyingcs@ruc.edu.cn}

\author{Teng Shi}\authornotemark[1]
\affiliation{%
  \institution{Gaoling School of AI,\\ Renmin University of
China}
    \city{Beijing}
  \country{China}
  }
\email{shiteng@ruc.edu.cn}

\author{Xiao Zhang}
\authornote{Xiao Zhang is the corresponding author.}
\affiliation{%
  \institution{Gaoling School of AI,\\ Renmin University of
China}
    \city{Beijing}
  \country{China}
  }
\email{zhangx89@ruc.edu.cn}

\author{Qi Liu}
\affiliation{%
  \institution{Wechat, Tencent}
  \city{Beijing}
  \country{China}}
\email{addisliu@tencent.com}

\author{Ruobing Xie}
\affiliation{%
  \institution{Wechat, Tencent}
  \city{Beijing}
  \country{China}}
\email{xrbsnowing@163.com}

\author{Jun Xu}
\affiliation{%
  \institution{Gaoling School of AI,\\ Renmin University of
China}
  \city{Beijing}
  \country{China}}
\email{junxu@ruc.edu.cn}

\author{Ji-Rong Wen}
\affiliation{%
  \institution{Gaoling School of AI,\\ Renmin University of
China}
  \city{Beijing}
  \country{China}}
\email{jrwen@ruc.edu.cn}

\renewcommand{\shortauthors}{Changshuo Zhang et al.}
\newcommand{\ourname}{$\text{Transition}^2$}
\input{section/0.abstract}



\maketitle

\input{section/1.intro}
\input{section/2.related}
\input{section/3.method}
\input{section/4.exp}
\input{section/5.conclusion}

\bibliographystyle{ACM-Reference-Format}
\bibliography{ref}


\end{document}

%% file: section/0.abstract.tex
\begin{abstract}

Nowadays, many recommender systems encompass various domains to cater to users' diverse needs, leading to user behaviors transitioning across different domains. 
In fact, user behaviors across different domains reveal changes in preference toward recommended items. For instance, a shift from negative feedback to positive feedback indicates improved user satisfaction.
However, existing cross-domain sequential recommendation methods typically model user interests by focusing solely on information about domain transitions, often overlooking the valuable insights provided by users' feedback transitions.
In this paper, we propose \ourname, a novel method to model transitions across both domains and types of user feedback. Specifically, \ourname~ introduces a transition-aware graph encoder based on user history, assigning different weights to edges according to the feedback type. This enables the graph encoder to extract historical embeddings that capture the transition information between different domains and feedback types.
Subsequently, we encode the user history using a cross-transition multi-head self-attention, incorporating various masks to distinguish different types of transitions. 
To further enhance representation learning, we employ contrastive losses to align transitions across domains and feedback types.
Finally, we integrate these modules to make predictions across different domains. Experimental results on two public datasets demonstrate the effectiveness of \ourname. 
\end{abstract}

%% file: section/1.intro.tex
\section{Introduction}
\begin{figure*}[t]
\centering
    \subfigure[An illustration of a user's transitions: 
    the first transition is a Type~1 transition, where the user enjoys a movie adaptation of a book, but negative (neg.) feedback arises from recommending the original book rather than its sequel; 
    the second transition belongs to Type~2, where the user is dissatisfied with art books and recommending other types of movies results in positive (pos.) feedback.]{
    \label{fig:intro_example}
    \includegraphics[width=1.2\columnwidth]{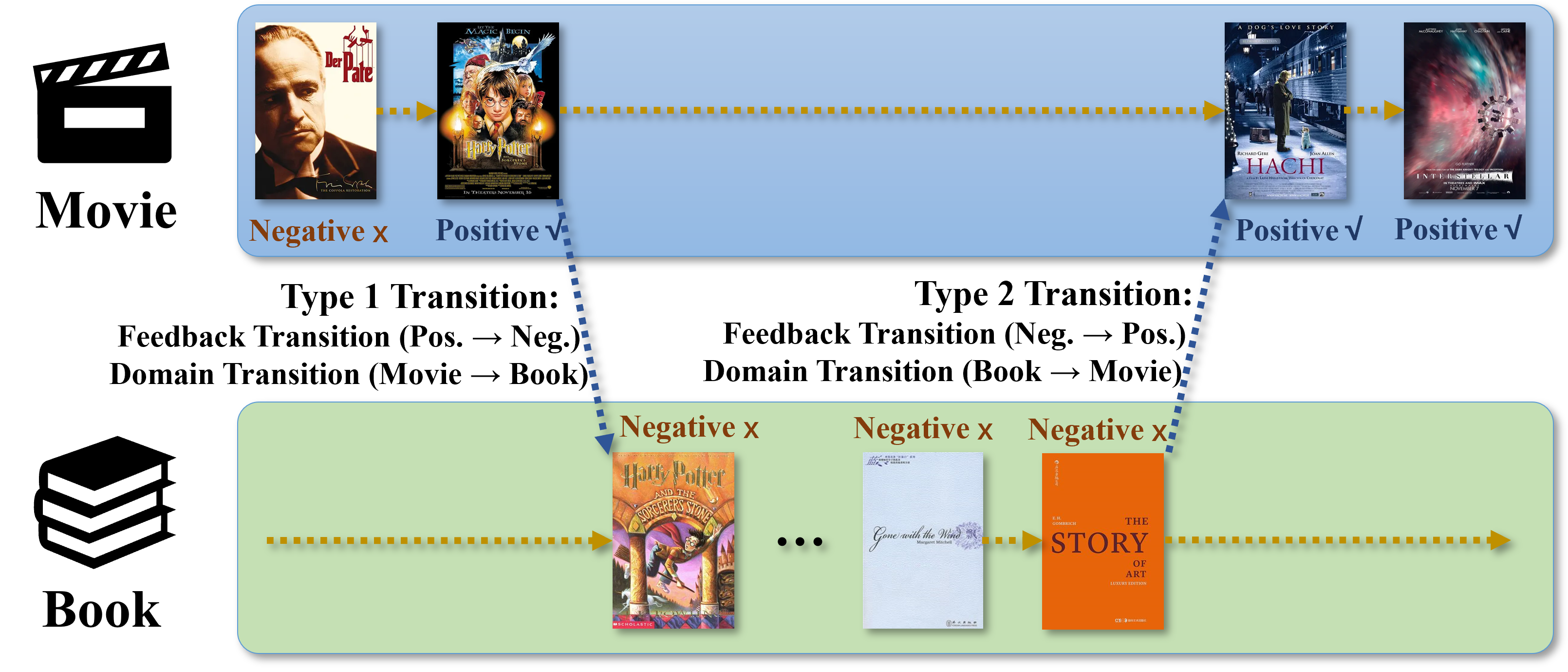}}
    \hspace{2mm}
    \subfigure[Statistical analyses on Book-Movie domains and Book-Music domains of Douban, focusing on the percentage of Type~1 transitions, Type~2 transitions, and other transitions among all cross-domain user behaviors.] {
    \label{fig:intro_val}
    \includegraphics[width=.8\columnwidth]{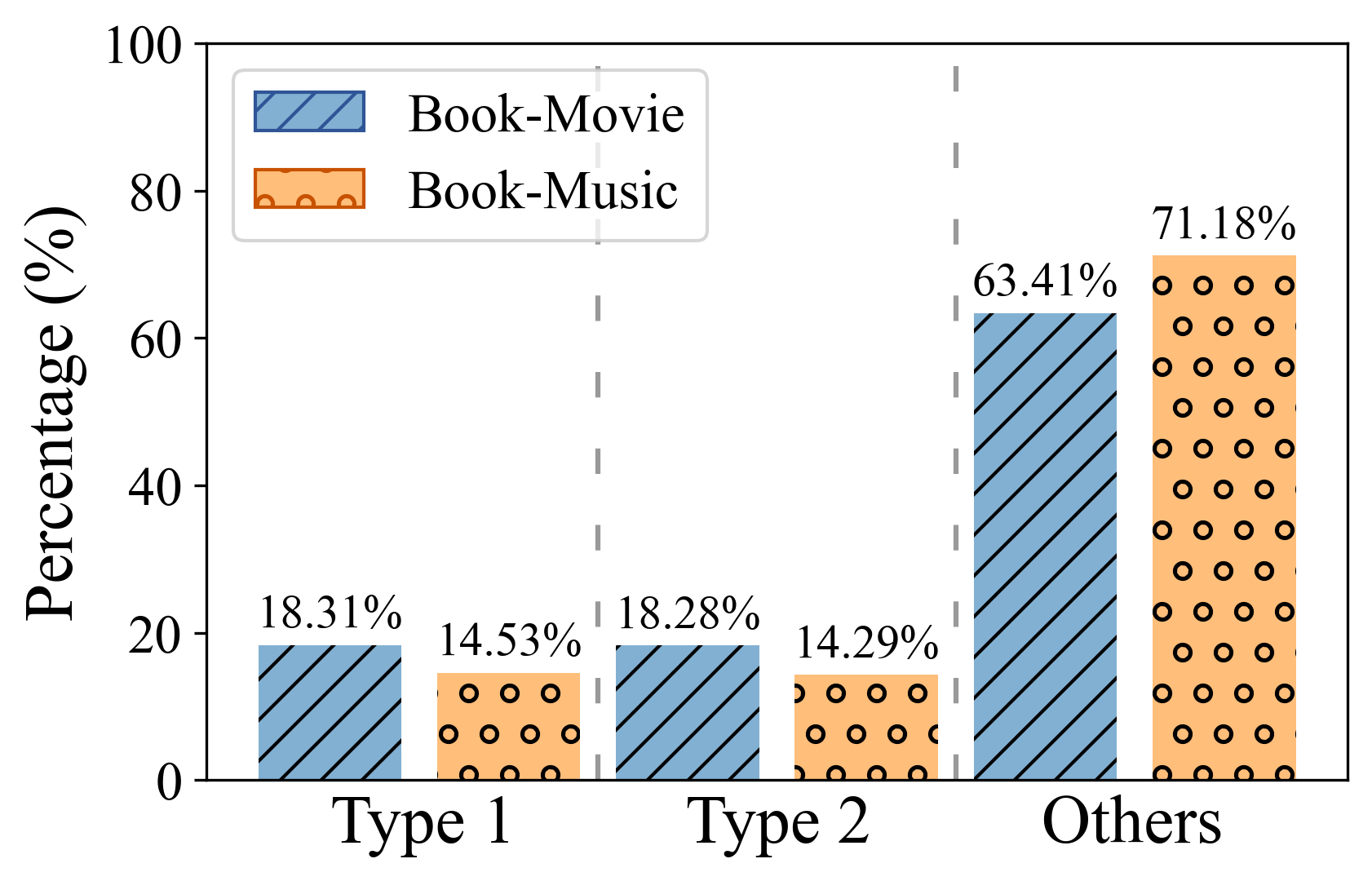}}
    \caption{A toy illustration and statistics of domain and feedback transitions in cross-domain sequential recommendation. Type~1 transition: when transitioning across domains, the user feedback changed from positive to negative; Type~2 transition: when the user browses the new domain, her feedback improved from negative to positive.}
\end{figure*}

Traditional sequential recommendation models~\cite{zhang2024reinforcing,kang2018self,zhang2025test, zhang2024saqrec, tang2025think, zheng2023reciprocal} a user's history within a single domain, such as music~\cite{zhang2022counteracting} or videos~\cite{zhao2024counteracting, zhang2025comment}. However, modern recommender systems often encompass multiple domains, such as books~\cite{anwar2019machine} and movies~\cite{goyani2020review}, leading users to transition between different domains. In this transition process, not only is the information about the domain transitions valuable (referred to  \emph{domain transition}), but the changes in user preference toward recommended items during domain transitions are also crucial (termed \emph{feedback transition}).

We classify users' feedback transitions during their domain transitions into two types:
(1)~\emph{Type~1 transition}: when transitioning across domains, the user feedback changed from positive to negative;  (2)~\emph{Type~2 transition}: when browsing the new domain, the user's feedback improves from negative to positive. 
For example, as shown in Figure~\ref{fig:intro_example}, for the transition of ``Type 1'', a user may search for related books after watching a movie she likes (positive). However, since the movie is adapted from a book, directly recommending the original book might be ignored (negative) because she has already watched the movie. In this case, the ideal recommendation would be the sequel to that book.
For a ``Type 2'' transition, the user is dissatisfied (negative) with several consecutively recommended books and turns to watch movies (positive) instead. Thus, recommending cross-domain items may benefit users consistently dissatisfied with one domain's items.

Furthermore, we analyzed the percentage of these two types of transitions among all transitions in the Douban\footnote{\url{https://www.douban.com/}} data, as shown in Figure~\ref{fig:intro_val}. Specifically, we analyzed the cross-domain behaviors between Book-Movie and Book-Music, focusing on the counts of ``Type 1'' and ``Type 2'' transitions. We found that these two transition types account for a significant percentage of all cross-domain behaviors ($18.31\%$ and $18.28\%$ in Book-Movie, and $14.53\%$ and $14.29\%$ in Book-Music). This finding underscores the importance of modeling cross-domain behaviors and feedback transitions.
This example shows that considering negative feedback can help us better understand users' cross-domain behavior transitions, enabling better user modeling.

Existing work typically focuses on modeling users' domain transition behaviors, 
to improve the performance of cross-domain sequential recommendations.
A pioneering work transfers single-domain representations learned from a single domain to other domains using a gated transfer module~\cite{ma2019pi}. Recent work models single-domain and cross-domain sequences separately and generates representations through a self-attention mechanism~\cite{cao2022contrastive,ma2024triple}. Despite the effectiveness of existing works, they overlooked the impact of users' feedback transitions during cross-domain behaviors, and often only focused on the positive feedback from users.

To model both the domain transitions and feedback transitions in users' cross-domain behaviors, this paper proposes an approach named \textbf{\ourname}, which models domain and feedback transitions for the cross-domain sequential recommendation. 
Firstly, we encode users' mixed histories across different domains using the transition-aware graph encoder. Specifically, when constructing the graph, we connect consecutive items in the history and assign different weights to the edges based on the feedback between the two items. 
This allows us to capture transitions between different domains and different types of feedback.
After encoding with the graph encoder, we obtain item embeddings that capture transitions between different domains and feedback types. 
We then input these history embeddings into a cross-transition multi-head self-attention to further model the transition information in the user's history.
To model the transitions between different domains and feedback types, we introduce different masks into the heads of the transformer's multi-head self-attention. This allows each head to capture different types of transitions. We then fuse the outputs from different heads to obtain the final history representations which include various transition information.
Furthermore, we design contrastive losses to align transitions across different domains and feedback types, helping the model better learn representations of different transitions.
Finally, we use these representations for prediction tasks in different domain recommendations.
The major contributions of the paper are as follows:
\begin{itemize}[leftmargin=*]
    \item We identified the importance of domain transitions and feedback transitions in cross-domain sequential recommendation and validated this through data statistics.
    \item We proposed \ourname, which models domain and feedback transitions using a GNN, a transformer equipped with different masks and a transition alignment module.
    \item Experiment results on two public datasets demonstrate the effectiveness of \ourname. \ourname~ outperforms existing sequential recommendation and cross-domain sequential recommendation models.
\end{itemize}

%% file: section/2.related.tex
\section{Related Work}
Cross-Domain Sequential Recommendation (CDSR) addresses the challenge of recommending items across diverse domains by leveraging various methodological approaches. Perera and Zimmerman pioneered using timestamp information to divide user interactions into temporal itemsets, enhancing the modeling of users' dynamic interests across domains~\cite{perera2020lstm,perera2020towards}. Zhang et al. further advanced this by proposing the CGN model, utilizing dual generators to map itemsets across domains concurrently~\cite{zhang2020learning}. Another strand of research focuses on domain-specific sequential modeling, exemplified by $\pi$-Net~\cite{ma2019pi} and DASL~\cite{li2021dual}, which employ GRUs and attention mechanisms for knowledge transfer. Recent innovations like DA-GCN~\cite{chen2019dagcn} and MIFN~\cite{ma2022mixed} introduce graph-based approaches to link domain-specific item sequences, while industry applications such as SEMI~\cite{lei2021semi} and RecGURU~\cite{li2022recguru} apply adversarial learning and multi-modal data fusion for cross-domain short-video recommendations. Novel hybrid models like C2DSR~\cite{cao2022contrastive} integrate graphical and attentional mechanisms, employing contrastive objectives to enhance both intra-domain and cross-domain user representations. DREAM~\cite{ye2023dream} focuses on modeling decoupled representations for both single- and cross-domain. TriCDR~\cite{ma2024triple} uses triple cross-domain attention and contrastive learning to model comprehensive cross-domain correlations. CA-CDSR~\cite{yin2024learning} focus on item representation alignment via sequence-aware augmentation and adaptive partial alignment.
In summary, current CDSR methods fail to effectively extract and integrate cross-domain transition information, disregarding user negative feedback and thus neglecting the transition of information between positive and negative feedback. These challenges are crucial in cross-domain recommendation scenarios for predicting future user interactions.

\begin{figure*}[t]
\centering
\includegraphics[width=1.0\textwidth]{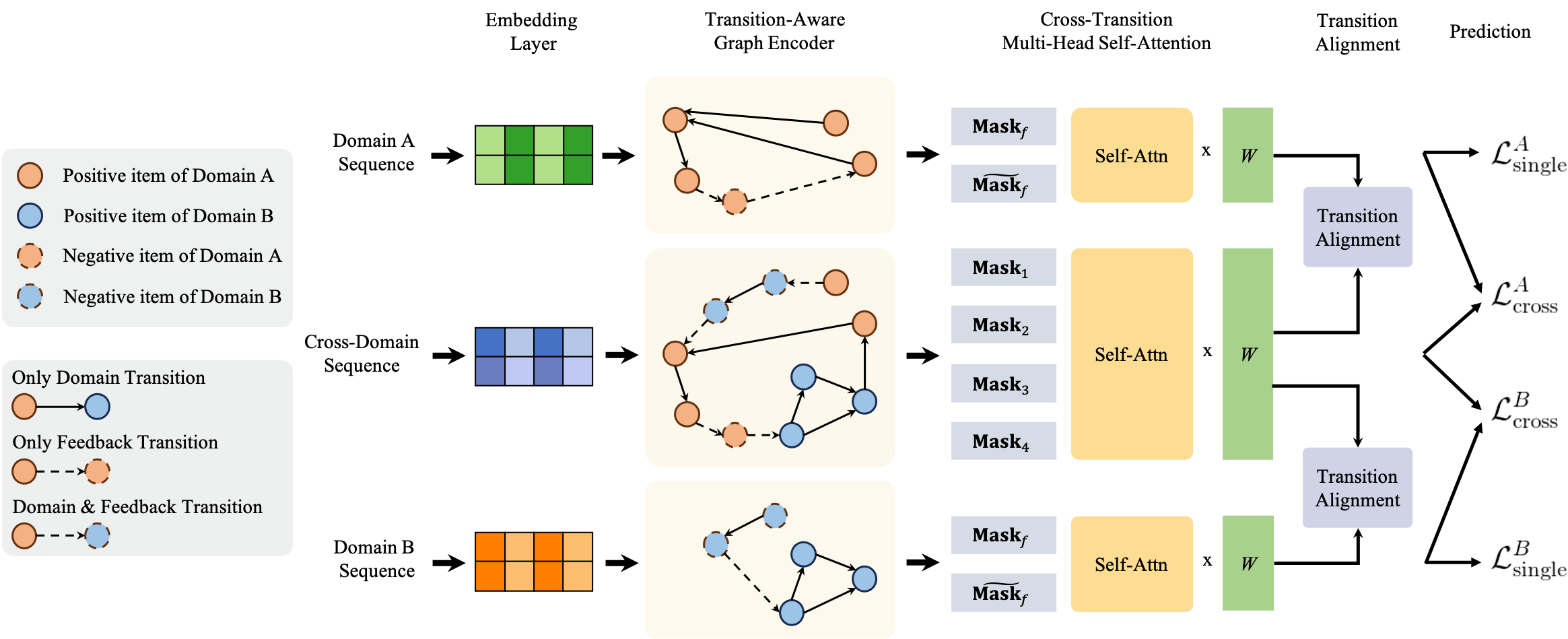}
\caption{
The overall architecture of \ourname~ comprises five key components: (1) an embedding layer, which initializes item embeddings; (2) a transition-aware graph encoder, which constructs graph-based representations of user sequences while incorporating transition information; (3) a cross-transition multi-head self-attention mechanism, designed to capture diverse transition patterns through specialized masking strategies; (4) a transition alignment module, which employs contrastive learning to align representations across different transitions; and (5) a prediction module, responsible for generating recommendations and optimizing the overall objective.
}
\label{main_model2}
\end{figure*}

%% file: section/3.method.tex
\section{Task Formulation}
In this paper, we discuss a comprehensive CDSR scenario where each user history involves two domains $A$ and $B$. Each instance $S_A, S_B, S_{C}$ corresponds to a specific user. For any given instance, $S_A=[(a_1, y_1), \ldots, (a_{|S_A|}, y_{|S_A|})]$ and $S_B = [(b_1, y'_1), \ldots, (b_{|S_B|}, y'_{|S_B|})]$ represent the single-domain user histories, where $y_i$ and $y'_i$ indicate the feedback associated with $a_i$ and $b_i$, respectively, taking values of $+1$ (positive feedback) or $-1$ (negative feedback). The sequence $S_{C} = [(b_1, y'_1), (a_1, y_1), \ldots, (b_{|S_B|}, y'_{|S_B|}), \ldots, (a_{|S_A|}, y_{|S_A|})]$ represents the cross-domain user history, formed by merging $S_A$ and $S_B$ in chronological order, where each $a \in A$ and $b \in B$ are the interacted items, and $|\cdot|$ denotes the total number of items. Note that $A$ and $B$ denote the entire item sets in domain $A$ and domain $B$, respectively. 
Given the observed user history $(S_A, S_B, S_{C})_u$, the goal of CDSR is to predict the next item:
\[
\arg\max_{a_i \in A} P_A(a_{|S_A|+1} | S_A, S_B, S_{C}) \quad \text{if the next item is in } A,
\]
\[
\arg\max_{b_j \in B} P_B(b_{|S_B|+1} | S_A, S_B, S_{C}) \quad \text{if the next item is in } B,
\]
where $P_A(a_i | S_A, S_B, S_{C}) \in \mathbb{R}^{|A|}$ and $P_B(b_j | S_A, S_B, S_{C}) \in \mathbb{R}^{|B|}$ denote the probability of the candidate item in domain $A$ and $B$, respectively, with the highest probability item being chosen as the next recommended item.

\section{Transition$^2$: The Proposed Algorithm}

In this section, we will introduce our proposed model~\ourname.
\subsection{Overview of Transition$^2$}
The overall framework of our model Transition$^2$ is illustrated in Figure~\ref{main_model2}. To effectively capture domain-transition and feedback-transition information: 
(1)~\ourname~ first initializes the embeddings for two single-domain sequences and the cross-domain sequence with three embedding layers. 
(2)~\ourname~ then constructs \emph{transition-aware graph encoder} based on users' cross-domain and single-domain sequences. Unlike previous methods that only utilize positive feedback for graph construction, we incorporate negative feedback and adjust the weight matrix of the graph neural network accordingly. 
(3)~After obtaining the representation of the user's historical sequences through the graph encoder, we design a \emph{cross-transition multi-head self-attention} mechanism. This mechanism calculates attention between and within different behaviors of the user's sequence to capture transition information.
(4)~A transition alignment module with two contrastive losses aligns intra-domain item representations while contrasting those with different feedback types.
(5)~The predicted scores are then calculated based on the final sequence representations, optimizing both single-domain and cross-domain losses.

\subsection{Transition-Aware Graph Encoder}
Inspired by the advantages of GNNs in handling sequential recommendations~\cite{wu2019session,xu2019graph}, we leverage GNNs to transform users' sequential behaviors into graph representations. Additionally, to effectively capture domain transitions and feedback transitions, we have introduced transition-aware graph encoder.
\subsubsection{Embedding Initialization}
For the sequential recommendation settings, we construct single-domain graphs $G_A$ and $G_B$, as well as a cross-domain graph $G_C$, based on all users' historical sequences, where only item nodes are included. Consequently, we initialize the item embeddings~\cite{qu2023continuous,qu2024budgeted} in the three graphs as $\mathbf{E}^{(0)}_{A}=[\mathbf{e}^{(0)}_{A1};\mathbf{e}^{(0)}_{A2};\cdots]\in \mathbb{R}^{{|A|}\times d}$, $\mathbf{E}^{(0)}_{B}=[\mathbf{e}^{(0)}_{B1};\mathbf{e}^{(0)}_{B2};\cdots]\in \mathbb{R}^{{|B|}\times d}$, $\mathbf{E}^{(0)}_{C}=[\mathbf{e}^{(0)}_{C1};\mathbf{e}^{(0)}_{C2};\cdots]\in \mathbb{R}^{{(|A|+|B|)}\times d}$, respectively.
\subsubsection{Transition-Aware Graph Construction}
The sequence graphs are constructed based on the user's interaction sequences by treating each item in the sequence as a node and adding edges between consecutive items. The complete graphs is obtained by combining all sequences in the dataset. Unlike previous methods that only consider items with positive feedback for graph construction, we also include items with negative feedback. The comprehensive cross-domain graph includes all items with both positive and negative feedback from domains A and B. However, this approach introduces some challenges. Firstly, during the propagation phase of the graph, nodes receive information from their neighbors. Essentially, items with different feedback should propagate opposite information. For example, items with positive feedback should receive positive information from neighboring items with positive feedback and negative information from neighboring items with negative feedback. Therefore, we incorporate feedback-transition information to obtain the transition matrix $\mathbf{T}$. If two adjacent nodes have different types of feedback, we set their corresponding values opposite. Specifically, for two adjacent items \(i\) and \(j\) with corresponding feedback \(y_i\) and \(y_j\), the transition matrix is defined as follows:
\begin{equation}
\textbf{T}_{ij} = \begin{cases} 
1 & \text{if } y_i = y_j, \\
-1 & \text{if } y_i \neq y_j, \\
0 & \text{if } i \text{ and } j \text{ are not adjacent}.
\end{cases}    
\end{equation}
We then obtain the transition-aware adjacency matrix of the item-item transition graph as
\begin{equation}
\mathbf{W}=\left(\begin{array}{cc}
\mathbf{0} & \mathbf{T} \\
\mathbf{T}^\top & \mathbf{0} \\
\end{array}\right).
\end{equation}
To stabilize training, we use the normalized form:
\begin{equation}
\mathbf{\widehat{W}} = \mathbf{D}^{-\frac{1}{2}}\mathbf{W}\mathbf{D}^{-\frac{1}{2}}.
\end{equation}
In this equation, \(\mathbf{D} \in \mathbb{R}^{(|A|+|B|) \times (|A|+|B|)}\) is a diagonal matrix where \(\mathbf{D}_{ii}\) represents the count of nonzero elements in the \(i\)-th row of \(\mathbf{W}\). Specifically, for single-domain graphs A and B and the cross-domain graph, we separately define their transition matrices as $\mathbf{\widehat{W}}_{A}$, $\mathbf{\widehat{W}}_{B}$, $\mathbf{\widehat{W}}_{C}$.

\subsubsection{Propagation and Aggregation}
For each graph, we adopt a widely used LightGCN-based propagation approach, which abandons feature transformation and non-linear activation, instead opting for simple weighted sum aggregation operators, with the $k$-th layer's propagation defined as
\begin{equation}
\mathbf{e}_i^{(k+1)}=\sum_{j \in \mathcal{N}_i} \frac{\textbf{T}_{ij}}{\sqrt{\left|\mathcal{N}_i\right|} \sqrt{\left|\mathcal{N}_j\right|}} \mathbf{e}_i^{(k)},
\end{equation}
where $\mathbf{e}^{(k)}_i$ is in $\mathbf{E}^{(k)}=[\mathbf{e}^{(k)}_{1};\mathbf{e}^{(k)}_{2};\cdots]$, and $\mathcal{N}_i$ and $\mathcal{N}_j$ respectively represent the number of neighbors for nodes $i$ and $j$ in the graph.
Then we can formulate the matrix equivalent form of the propagation for the three graphs as
\begin{equation}
\label{eq:prop}
\left\{\begin{array}{l}
    \mathbf{E}^{(k+1)}_A = \mathbf{\widehat{W}}_A \mathbf{E}^{(k)}_A,\\ \mathbf{E}^{(k+1)}_B = \mathbf{\widehat{W}}_B \mathbf{E}^{(k)}_B,\\
\mathbf{E}^{(k+1)}_{C} = \mathbf{\widehat{W}}_{C} \mathbf{E}^{(k)}_{C}.\\
\end{array}\right.
\end{equation}
Finally, after passing through $K$ layers, the final embeddings of the three graphs are obtained by:
\begin{equation}
\label{eq:final_node}
\left\{\begin{array}{l}
\mathbf{E}_A =\frac{1}{K} \sum_{k=0}^K \mathbf{E}^{(k)}_A,\\ \mathbf{E}_B =\frac{1}{K} \sum_{k=0}^K \mathbf{E}^{(k)}_B,\\
\mathbf{E}_{C} =\frac{1}{K} \sum_{k=0}^K \mathbf{E}^{(k)}_{C}.\\
\end{array}\right.
\end{equation}

\subsection{Cross-Transition Multi-Head Self-Attention}
In this section, we introduce the Cross-Transition Multi-Head Self-Attention module in \ourname~ that captures cross-domain transition and feedback transition information by equipping the multi-head self-attention module with a cross-mask mechanism.

Firstly, to model domain transitions and feedback transitions separately in the cross-domain user history, given the user's ordered interaction history \(\{ \text{item}_1, \text{item}_2, \ldots \}\) with corresponding domain identifiers \(\{ d_1, d_2, \ldots \}\) and feedbacks \(\{ f_1, f_2, \ldots \}\), we first obtain the cross-domain embedded representation $\mathbf{E}_{C}$ through the transition-aware graph encoder for this sequence. Next, to capture sequential relationships, we further introduce positional embedding $\mathbf{P}_C$. For the self-attention mechanism, we define $\widehat{\mathbf{E}}_{C}=\mathbf{E}_{C}+\mathbf{P}_C$ as the input. However, simply using all embeddings as input without any restrictions would make it difficult for the self-attention mechanism to distinguish between feedback and domain information. Therefore, we introduce two masks $\mathbf{M}^{f}$ and $\mathbf{M}^{d}$ to better capture feedback transitions and domain transitions, with each entry defined as:

\begin{equation}
\mathbf{M}^{f}_{i j}= \begin{cases}0 & \text { if } f_i=f_j, \\ 1 & \text { if } f_i \neq f_j.\end{cases}
\end{equation}

\begin{equation}
\mathbf{M}^{d}_{i j}= \begin{cases}0  & \text { if } d_i=d_j, \\ 1 & \text { if } d_i \neq d_j.\end{cases}
\end{equation}

However, simply introducing these two masking mechanisms is not sufficient to strongly capture cross-transition information. Therefore, we perform a cross-processing of these two masks to obtain cross-masks that can capture four types of cross-transition information:
\begin{equation}
\label{eq:final_mask}
\left\{\begin{array}{l}
\mathbf{M}_1 =\mathbf{M}^{f} \& \mathbf{M}^{d},\\ \mathbf{M}_2 =\mathbf{M}^{f} \& \widetilde{\mathbf{M}}^{d},\\
\mathbf{M}_3 =\widetilde{\mathbf{M}}^{f} \& \mathbf{M}^{d},\\
\mathbf{M}_4 = \widetilde{\mathbf{M}}^{f} \& \widetilde{\mathbf{M}}^{d}.
\end{array}\right.
\end{equation}
where \& represents bitwise AND, and $\widetilde{\mathbf{M}}$ represents the negation of $\mathbf{M}$. To account for all four types of cross-transitions, we equip each head of the Multi-Head Self-attention with a different mask. Specifically, we set the overall mask as $\mathbf{M}_{C} = [\mathbf{M}_1; \mathbf{M}_2; \mathbf{M}_3; \mathbf{M}_4; \cdots]$. The Cross-Transition MHSA then takes $\widehat{\mathbf{E}}_{C}$ and $\mathbf{M}$ as inputs, with $\mathbf{Q}=\mathbf{K}=\mathbf{V}=\widehat{\mathbf{E}}_{C}$, and a Feed-Forward layer following, and output the final embedding $\mathbf{E}^\prime_{C}$:
\begin{equation}
\mathbf{E}^\prime_{C}=\mathrm{FFN}\left(\mathrm{MHSA}\left(\widehat{\mathbf{E}}_{C}, \widehat{\mathbf{E}}_{C}, \widehat{\mathbf{E}}_{C}, \mathbf{M}_{C}\right)\right).
\end{equation}
Specifically, the detailed computation of Multi-Head Self-Attention is as follows:
\begin{equation}
    \mathrm{MHSA}\left(\widehat{\mathbf{E}}_{C}, \widehat{\mathbf{E}}_{C}, \widehat{\mathbf{E}}_{C}, \mathbf{M}_{C}\right)=\text {Concate}\left(\text {head}_{1}, \text {head}_{2}, \ldots\right) W.
\end{equation}
where for each head$_{i=1,2,\ldots}$,
\begin{equation}
\begin{aligned}    \text{head}_{i}&=\mathrm{Attention}\left(\widehat{\mathbf{E}}_{C},\widehat{\mathbf{E}}_{C},\widehat{\mathbf{E}}_{C}, \mathbf{M}_{i\%4}\right)\\
    & =\mathrm{Softmax}\left(\widehat{\mathbf{E}}_{C}\widehat{\mathbf{E}}_{C}^{\top} / \sqrt{d / h} \odot \mathbf{M}_{i\%4}\right) \widehat{\mathbf{E}}_{C},
\end{aligned}
\end{equation}
where $\odot$ denotes the Hadamard product.

However, for single-domain sequences, domain transitions do not occur, so we only consider feedback transitions. Specifically, for the single-domain embedded representations \(\mathbf{E}_A \) and \( \mathbf{E}_B \) after the Transition-Aware Graph encoder and the positional embedding $\mathbf{P}_{A}$ and $\mathbf{P}_{B}$, we define $\widehat{\mathbf{E}}_{A}=\mathbf{E}_{A}+\mathbf{P}_{A}$, $\widehat{\mathbf{E}}_{B}=\mathbf{E}_{B}+\mathbf{P}_{B}$, \( \mathbf{M}_A = \mathbf{M}_B = [\mathbf{M}^f; \widetilde{\mathbf{M}}^f; \mathbf{M}^f; \cdots] \). Then, the final embedded representations are obtained through a similar Multi-Head Self-Attention mechanism:
\begin{equation}
\mathbf{E}^\prime_{A}=\mathrm{FFN}\left(\mathrm{MHSA}\left(\widehat{\mathbf{E}}_{A}, \widehat{\mathbf{E}}_{A}, \widehat{\mathbf{E}}_{A}, \mathbf{M}_{A}\right)\right),
\end{equation}
\begin{equation}
\mathbf{E}^\prime_{B}=\mathrm{FFN}\left(\mathrm{MHSA}\left(\widehat{\mathbf{E}}_{B}, \widehat{\mathbf{E}}_{B}, \widehat{\mathbf{E}}_{B}, \mathbf{M}_{B}\right)\right).
\end{equation}

\subsection{Transition Alignment and Contrast}
\label{cl}

For the extracted three types of representations, where $\mathbf{E}^\prime_{A}$ represents the characterization of domain A, $\mathbf{E}^\prime_{B}$ represents domain B, and $\mathbf{E}^\prime_{C}$ denotes the mix-domain representations, we first align the item representations within the same domain across these three sets, and then conduct contrastive modeling on representations with different feedback types to capture their nuanced differences.

\subsubsection{Domain Representation Alignment}
Given that we have three sets of representation information, the goal is to guide the model to pull semantically equivalent representations closer within the same domain. To achieve this, we introduce InfoNCE~\cite{oord2018representation} to align the domain A information in $\mathbf{E}^\prime_{C}$ with $\mathbf{E}^\prime_{A}$, and the domain B information in $\mathbf{E}^\prime_{C}$ with $\mathbf{E}^\prime_{B}$. Specifically, for $\mathbf{E}^\prime_{C}$, we separate domain A and domain B historical records in each user sequence through indexing to obtain $\mathbf{E}^{\prime\prime}_{A}$ and $\mathbf{E}^{\prime\prime}_{B}$. We then apply mean-pooling along the sequence dimension to derive:
$\mathbf{e}^{\prime\prime}_{A} = \text{MEAN}(\mathbf{E}^{\prime\prime}_{A})$ and 
$\mathbf{e}^{\prime\prime}_{B} = \text{MEAN}(\mathbf{E}^{\prime\prime}_{B})$.
Similarly, we perform mean-pooling for $\mathbf{E}^\prime_{A}$ and $\mathbf{E}^\prime_{B}$ with
$\mathbf{e}^\prime_{A} = \text{MEAN}(\mathbf{E}^\prime_{A})$ and $\mathbf{e}^\prime_{B} = \text{MEAN}(\mathbf{E}^\prime_{B})$.
The alignment loss is defined using in-batch InfoNCE loss:
\begin{small}
\begin{equation}
\begin{aligned}
\mathcal{L}_{\text{A-align}} = 
- \left[ \log \frac{\exp\left(\mathbf{e}^{\prime\prime\top}_{A} \mathbf{e}^{\prime}_A / \tau\right)}{\sum_{\mathbf{e}^{\prime-}_{A}\in\mathcal{N}_A^{\prime}} \exp\left(\mathbf{e}^{\prime\prime\top}_{A} \mathbf{e}^{\prime-}_{A} / \tau\right)} 
+ \log \frac{\exp\left(\mathbf{e}^{\prime\top}_{A} \mathbf{e}^{\prime\prime}_A / \tau\right)}{\sum_{\mathbf{e}^{\prime\prime-}_{A}\in\mathcal{N}_A^{\prime\prime}} \exp\left(\mathbf{e}^{\prime\top}_{A} \mathbf{e}^{\prime\prime-}_{A} / \tau\right)} \right],\\
\end{aligned}
\end{equation}
\end{small}
where $\tau$ is the temperature hyperparameter, $\mathcal{N}_A^{\prime}$ denotes other users’ histories within the same batch of $\mathbf{e}^{\prime}_A$ as negative samples, $\mathcal{N}_A^{\prime\prime}$ denotes other users’ histories within the same batch of $\mathbf{e}^{\prime\prime}_A$ as negative samples.
Similarly, we obtain the loss $\mathcal{L}_{\text{B-align}}$ for domain B.
The total alignment loss is as follows:
\begin{equation}
    \mathcal{L}_{\text{align}} = \mathcal{L}_{\text{A-align}} + \mathcal{L}_{\text{B-align}}.
\end{equation}

\subsubsection{Feedback Representation Contrast}
For items with different feedback types, we enforce their representations to be pulled apart within the same sequence to enable the model to perceive their differences. Specifically, we treat pairs of positive-feedback items in each user sequence as positive samples, and negative-feedback items as negative samples. For a general representation \(\mathbf{E}\) (applicable to \(\mathbf{E}^\prime_{A}\), \(\mathbf{E}^\prime_{B}\), and \(\mathbf{E}^\prime_{C}\)), we formulate the in-sequence triplet loss as follows:
\begin{equation}
\begin{aligned}
\mathcal{L}_{\text{cont}}^{\mathbf{E}} = \sum_{i=1}^{|\mathcal{P}_{+}|} \max\Big(   \|\mathbf{e}_{i} - \frac{1}{|\mathcal{P}_+|}\sum_{j=1}^{|\mathcal{P}_+|}\mathbf{e}_{j}\|_2^2 \\
- \|\mathbf{e}_{i} - \frac{1}{|\mathcal{N}_-|}\sum_{k=1}^{|\mathcal{N}_-|}\mathbf{e}_{k}\|_2^2 + \alpha, 0 \Big),
\end{aligned}
\end{equation}
where $\mathbf{e}_{i}$ represents the embedding vector of the $i$-th item in the sequence, \(\mathcal{P}_+\) denotes the set of positive sample indexes and \(\mathcal{N}_-\) denotes the set of negative sample indexes within the sequence. We compute this loss for \(\mathbf{E}^\prime_{A}\), \(\mathbf{E}^\prime_{B}\), and \(\mathbf{E}^\prime_{C}\) respectively, and sum them up to obtain the contrastive loss \(\mathcal{L}_{\text{cont}}\):
\begin{equation}
\mathcal{L}_{\text{cont}} = \mathcal{L}_{\text{cont}}^{\mathbf{E}^\prime_A} + \mathcal{L}_{\text{cont}}^{\mathbf{E}^\prime_B} + \mathcal{L}_{\text{cont}}^{\mathbf{E}^\prime_C}.
\end{equation}
By introducing the losses $\mathcal{L}_{\text{align}}$ and $\mathcal{L}_{\text{cont}}$, \ourname~ aligns these three sets of item representations and captures the differences among items with different feedback types within each user behavior sequence.
\subsection{Model Training}
For the final model training loss, we designed losses $\mathcal{L}_{\text {single }}^{A}$ and $\mathcal{L}_{\text {single }}^{B}$ for single-domain recommendation tasks and losses $\mathcal{L}_{\text {cross }}^{A}$ and $\mathcal{L}_{\text {cross }}^{B}$ for cross-domain recommendation tasks. These are then combined to obtain the total loss with $\mathcal{L}_{\text{align}}$ and $\mathcal{L}_{\text{cont}}$:
\begin{equation}
    \mathcal{L}_{\text {total }}=\underbrace{\mathcal{L}_{\text {single }}^{A}+\mathcal{L}_{\text {single }}^{B}}_{\text {Single-Domain Loss }}+\underbrace{\mathcal{L}_{\text {cross }}^{A}+\mathcal{L}_{\text {cross }}^{B}}_{\text {Cross-Domain Loss }}+\mu_1\mathcal{L}_{\text{align}}+\mu_2\mathcal{L}_{\text{cont}},
\end{equation}
where $\mu_1$ and $\mu_2$ are hyper-parameters and the recommendation losses are all defined as the cross-entropy loss between the predictions and the ground truth:
\begin{equation}
\label{eq:final_node}
\left\{\begin{array}{l}
\mathcal{L}_{\text {single }}^{A} =-\operatorname{log}\mathrm{Softmax}\left(\mathrm{MLP}_A(\mathbf{E}^\prime_{C}+\mathbf{E}^\prime_{A})\right)_{a_{|S_A|+1}},\\ 
\mathcal{L}_{\text {single }}^{B} =-\operatorname{log}\mathrm{Softmax}\left(\mathrm{MLP}_B(\mathbf{E}^\prime_{C}+\mathbf{E}^\prime_{B})\right)_{b_{|S_B|+1}},\\
\mathcal{L}_{\text {cross }}^{A} =-\operatorname{log}\mathrm{Softmax}\left(\mathrm{MLP}_A(\mathbf{E}^\prime_{C})\right)_{a_{|S_A|+1}},\\
\mathcal{L}_{\text {cross }}^{B} = -\operatorname{log}\mathrm{Softmax}\left(\mathrm{MLP}_B(\mathbf{E}^\prime_{C})\right)_{b_{|S_B|+1}}.
\end{array}\right.
\end{equation}

%% file: section/4.exp.tex
\section{Experiments}
We conduct extensive experiments and detailed studies to evaluate the performance of \ourname.


\begin{table}[t]
    \caption{
    Statistics of three domains on Douban.}
    \vspace{-8px}
    \center
     \resizebox{.95\columnwidth}{!}{
        \begin{tabular}{lccccc}
        \toprule
        Domain & \#Users & \#Items & \#Records & Density &Avg. Rating  \\
        \midrule
        Book &26,877 &187,520 &1,097,148 &0.0218\% &4.0391 \\
        Movie &28,718 &57,424 &2,828,585 &0.1715\% &3.8101 \\
        Music &23,822 &185,574 &1,387,216 &0.0314\% &4.1749 \\ 
        \bottomrule
        \end{tabular}}
    \label{tab:stat}
   \vspace{-0.3cm}
\end{table}


\begin{table}[t]
    \caption{
    Statistics of two domains on Amazon.}
    \vspace{-8px}
    \center
     \resizebox{.95\columnwidth}{!}{
        \begin{tabular}{lccccc}
        \toprule
        Domain & \#Users & \#Items & \#Records & Density &Avg. Rating  \\
        \midrule
        Book &18,788 &93,041 &22,507,155 &1.2876\%  &4.2958 \\
        Movie &12,807 &29,243 &4,607,046 &1.2301\% &4.1869\\
        \bottomrule
        \end{tabular}} 
    \label{tab:stat2}
   \vspace{-0.3cm}
\end{table}

\begin{table*}
    \centering
    \caption{Experimental results of the Book-Movie domains on Douban. The best result is bolded and the runner-up is underlined. Improvements over the second-best methods are significant (\textit{t}-test, \textit{p}-value $<$ 0.05).}
    \setlength\tabcolsep{9pt} 
\begin{tabular}{l|c|cc|ccc|c|cc|ccc}
\hline \multirow{3}{*}{ Methods } & \multicolumn{6}{c|}{ Book-domain recommendation } & \multicolumn{6}{c}{ Movie-domain recommendation } \\
\cline{2-13} & MRR & \multicolumn{2}{|c|}{$\mathrm{NDCG}$} & \multicolumn{3}{c|}{$\mathrm{HR}$} & MRR & \multicolumn{2}{|c|}{ NDCG } & \multicolumn{3}{c}{$\mathrm{HR}$} \\
\cline{2-13} & $@10$ & $@ 5$ & $@10$ & $@ 1$ & $@5$ & $@10$ & $@10$ & $@ 5$ & $@10$ & $@1$ & $@5$ & $@10$ \\
\hline
GRU4Rec & $0.131$ & $0.128$ & $0.139$ & $0.093$ & $0.159$ & $0.197$ & $0.162$ & $0.158$ & $0.180$ & $0.099$ & $0.213$ & $0.283$ \\
SASRec & $0.135$ & $0.132$ & $0.143$ & $0.098$ & $0.162$ & $0.197$ & $0.189$ & $0.187$ & $0.209$ & $0.124$ & $0.245$ & $0.313$ \\
SRGNN & $0.140$ & $0.138$ & $0.148$ & $0.100$ & $0.170$ & $0.200$ & $0.169$ & $0.163$ & $0.184$ & $0.111$ & $0.212$ & $0.278$ \\
\hline
CoNet & $0.135$ & $0.135$ & $0.148$ & $0.096$ & $0.172$ & $0.211$ & $0.162$ & $0.155$ & $0.179$ & $0.101$ & $0.207$ & $0.283$ \\
$\pi$-Net & $0.142$ & $0.144$ & $0.161$ & $0.091$ & $0.193$ & $0.248$ & $0.162$ & $0.158$ & $0.185$ & $0.090$ & $0.222$ & $0.308$ \\
C2DSR & $0.152$ & $0.149$ & $0.164$ & $0.106$ & $0.189$ & $0.245$ & $0.193$ & $0.192$ & $0.215$ & $0.125$ & $0.254$ & $0.326$ \\
TriCDR & $0.155$ & $0.154$ & $0.168$ & $0.110$ & $0.195$ & $0.239$ & $0.200$ & $0.198$ & $0.223$ & $0.130$ & $0.262$ & $0.337$\\
CA-CDSR & $\underline{0.161}$ & $\underline{0.163}$ & $\underline{0.180}$ & $\underline{0.115}$ & $\underline{0.205}$ & $\underline{0.259}$ & $\underline{0.214}$ & $\underline{0.210}$ & $\underline{0.235}$ & $\underline{0.143}$ & $\underline{0.279}$ & $\underline{0.359}$\\
\hline
\ourname & $\textbf{0.179}$ & $\textbf{0.178}$ & $\textbf{0.196}$ & $\textbf{0.124}$ & $\textbf{0.227}$ & $\textbf{0.283}$ & $\textbf{0.227}$ & $\textbf{0.227}$ & $\textbf{0.254}$ & $\textbf{0.150}$ & $\textbf{0.298}$ & $\textbf{0.379}$\\
\hline
\end{tabular}
    \label{tab:exp1}
\end{table*}

\begin{table*}
\caption{Experimental results of the Book-Music domains on Douban. The best result is bolded and the runner-up is underlined. Improvements over the second-best methods are significant (\textit{t}-test, \textit{p}-value $<$ 0.05).}
    \centering
    \setlength\tabcolsep{9pt} 
\begin{tabular}{l|c|cc|ccc|c|cc|ccc}
\hline \multirow{3}{*}{ Methods } & \multicolumn{6}{c|}{ Book-domain recommendation } & \multicolumn{6}{c}{ Music-domain recommendation } \\
\cline{2-13} & MRR & \multicolumn{2}{|c|}{$\mathrm{NDCG}$} & \multicolumn{3}{c|}{$\mathrm{HR}$} & MRR & \multicolumn{2}{|c|}{ NDCG } & \multicolumn{3}{c}{$\mathrm{HR}$} \\
\cline{2-13} & $@10$ & $@ 5$ & $@10$ & $@ 1$ & $@5$ & $@10$ & $@10$ & $@ 5$ & $@10$ & $@1$ & $@5$ & $@10$ \\
\hline
GRU4Rec & $0.124$ & $0.119$ & $0.131$ & $0.090$ & $0.147$ & $0.184$ & $0.118$ & $0.115$ & $0.129$ & $0.077$ & $0.151$ & $0.193$ \\
SASRec & $0.122$ & $0.120$ & $0.130$ & $0.088$ & $0.149$ & $0.182$ & $0.125$ & $0.123$ & $0.136$ & $0.084$ & $0.159$ & $0.198$ \\
SRGNN & $0.124$ & $0.122$ & $0.134$ & $0.091$ & $0.152$ & $0.186$ & $0.128$ & $0.125$ & $0.139$ & $0.086$ & $0.161$ & $0.205$ \\
\hline
CoNet & $0.130$ & $0.131$ & $0.144$ & $0.090$ & $0.169$ & $0.211$ & $0.123$ & $0.121$ & $0.139$ & $0.077$ & $0.164$ & $0.218$ \\
$\pi$-Net & $0.127$ & $0.128$ & $0.147$ & $0.075$ & $0.164$ & $0.204$ & $0.124$ & $0.121$ & $0.140$ & $0.072$ & $0.167$ & $0.210$ \\
C2DSR & $0.130$ & $0.126$ & $0.138$ & $0.094$ & $0.156$ & $0.192$ & $0.130$ & $0.130$ & $0.143$ & $0.088$ & $0.169$ & $0.210$ \\
TriCDR & $0.134$ & $0.131$ & $0.143$ & $0.095$ & $0.164$ & $0.203$ & $0.130$ & $0.128$ & $0.142$ & $0.086$ & $0.167$ & $0.210$ \\
CA-CDSR & $\underline{0.140}$ & $\underline{0.136}$ & $\underline{0.151}$ & $\underline{0.097}$ & $\underline{0.177}$ & $\underline{0.225}$ & $\underline{0.134}$ & $\underline{0.134}$ & $\underline{0.147}$ & $\underline{0.088}$ & $\underline{0.175}$ & $\underline{0.223}$ \\
\hline
\ourname & $\textbf{0.147}$ & $\textbf{0.144}$ & $\textbf{0.160}$ & $\textbf{0.099}$ & $\textbf{0.186}$ & $\textbf{0.235}$ & $\textbf{0.141}$ & $\textbf{0.138}$ & $\textbf{0.155}$ & $\textbf{0.090}$ & $\textbf{0.183}$ & $\textbf{0.235}$ \\
\hline
\end{tabular}
    \label{tab:exp2}
\end{table*}

\begin{table*}
\caption{Experimental results of the Book-Movie domains on Amazon. The best result is bolded and the runner-up is underlined. Improvements over the second-best methods are significant (\textit{t}-test, \textit{p}-value $<$ 0.05).}
    \centering
    \setlength\tabcolsep{9pt} 
\begin{tabular}{l|c|cc|ccc|c|cc|ccc}
\hline \multirow{3}{*}{ Methods } & \multicolumn{6}{c|}{ Book-domain recommendation } & \multicolumn{6}{c}{ Movie-domain recommendation } \\
\cline{2-13} & MRR & \multicolumn{2}{|c|}{$\mathrm{NDCG}$} & \multicolumn{3}{c|}{$\mathrm{HR}$} & MRR & \multicolumn{2}{|c|}{ NDCG } & \multicolumn{3}{c}{$\mathrm{HR}$} \\
\cline{2-13} & $@10$ & $@ 5$ & $@10$ & $@ 1$ & $@5$ & $@10$ & $@10$ & $@ 5$ & $@10$ & $@1$ & $@5$ & $@10$ \\
\hline
GRU4Rec & $0.085$ & $0.082$ & $0.085$ & $0.078$ & $0.080$ & $0.095$ & $0.101$ & $0.099$ & $0.105$ & $0.065$ & $0.131$ & $0.163$ \\
SASRec & $0.081$ & $0.079$ & $0.082$ & $0.073$ & $0.085$ & $0.092$ & $0.107$ & $0.103$ & $0.116$ & $0.071$ & $0.135$ & $0.172$ \\
SRGNN & $0.088$ & $0.085$ & $0.088$ & $0.085$ & $0.102$ & $0.116$ & $0.110$ & $0.115$ & $0.129$ & $0.075$ & $0.131$ & $0.175$ \\
\hline
CoNet & $0.092$ & $0.088$ & $0.095$ & $0.070$ & $0.105$ & $0.127$ & $0.101$ & $0.095$ & $0.110$ & $0.063$ & $0.125$ & $0.171$ \\
$\pi$-Net & $0.095$ & $0.092$ & $0.100$ & $0.069$ & $0.113$ & $0.129$ & $0.107$ & $0.100$ & $0.116$ & $0.068$ & $0.131$ & $0.179$ \\
C2DSR & $0.089$ & $0.086$ & $0.092$ & $0.075$ & $0.097$ & $0.114$ & $0.112$ & $0.107$ & $0.121$ & $0.077$ & $0.136$ & $0.177$\\
TriCDR & $\underline{0.098}$ & $\underline{0.094}$ & $0.099$ & $0.083$ & $0.104$ & $0.119$ & $0.121$ & $0.115$ & $0.124$ & $\underline{0.095}$ & $0.134$ & $0.163$ \\
CA-CDSR & $0.094$ & $0.093$ & $\underline{0.102}$ & $\underline{0.085}$ & $\underline{0.114}$ & $\underline{0.129}$ & $\underline{0.124}$ & $\underline{0.123}$ & $\underline{0.135}$ & $0.087$ & $\underline{0.143}$ & $\underline{0.179}$ \\
\hline
\ourname & $\textbf{0.106}$ & $\textbf{0.104}$ & $\textbf{0.120}$ & $\textbf{0.094}$ & $\textbf{0.113}$ & $\textbf{0.139}$ & $\textbf{0.136}$ & $\textbf{0.134}$ & $\textbf{0.145}$ & $\textbf{0.102}$ & $\textbf{0.161}$ & $\textbf{0.194}$ \\
\hline
\end{tabular}
    \label{tab:exp3}
\end{table*}

\begin{table*}[ht]
\caption{Ablation studies on the Book-Movie Domains on Douban.}
    \centering
    \setlength\tabcolsep{9pt} 
\begin{tabular}{l|c|cc|ccc|c|cc|ccc}
\hline \multirow{3}{*}{ Methods } & \multicolumn{6}{c|}{ Book-domain recommendation } & \multicolumn{6}{c}{ Movie-domain recommendation } \\
\cline{2-13} & MRR & \multicolumn{2}{c|}{$\mathrm{NDCG}$} & \multicolumn{3}{c|}{$\mathrm{HR}$} & MRR & \multicolumn{2}{c|}{ NDCG } & \multicolumn{3}{c}{$\mathrm{HR}$} \\
\cline{2-13} & $@10$ & $@5$ & $@10$ & $@1$ & $@5$ & $@10$ & $@10$ & $@ 5$ & $@10$ & $@1$ & $@5$ & $@10$ \\
\hline 
\hline 
w/o TAG & $0.163$ & $0.162$ & $0.175$ & $0.110$ & $0.207$ & $0.260$ & $0.213$ & $0.214$ & $0.230$ & $0.140$ & $0.281$ & $0.259$ \\
w/o CTM & $0.161$ & $0.161$ & $0.171$ & $0.108$ & $0.197$ & $0.249$ & $0.209$ & $0.208$ & $0.222$ & $0.135$ & $0.274$ & $0.252$ \\
w/o $\mathcal{L}_{\text{align}}$ & $0.165$ & $0.164$ & $0.176$ & $0.111$ & $0.205$ & $0.256$ & $0.210$ & $0.211$ & $0.218$ & $0.141$ & $0.285$ & $0.265$ \\
w/o $\mathcal{L}_{\text{cont}}$ & $0.170$ & $0.170$ & $0.186$ & $0.113$ & $0.215$ & $0.268$ & $0.217$ & $0.218$ & $0.232$ & $0.143$ & $0.287$ & $0.262$ \\
\hline
only $\mathbf{M}_1$ & $\underline{0.176}$ & $\underline{0.176}$ & $\underline{0.191}$ & $\underline{0.122}$ & $0.223$ & $\underline{0.273}$ & $\underline{0.226}$ & $\underline{0.225}$ & $\underline{0.252}$ & $\underline{0.146}$ & $\underline{0.291}$ & $0.372$ \\
only $\mathbf{M}_2$ & $0.172$ & $0.172$ & $0.185$ & $0.116$ & $0.213$ & $0.266$ & $0.221$ & $0.221$ & $0.247$ & $0.142$ & $0.290$ & $0.369$ \\
only $\mathbf{M}_3$ & $0.172$ & $0.172$ & $0.190$ & $0.120$ & $0.221$ & $0.270$ & $0.221$ & $0.221$ & $0.248$ & $0.144$ & $0.292$ & $0.370$ \\
only $\mathbf{M}_4$ & $0.175$ & $0.174$ & $0.182$ & $0.117$ & $0.214$ & $0.269$ & $0.224$ & $0.223$ & $0.249$ & $0.143$ & $0.289$ & $0.374$ \\
\hline
only $\mathbf{M}_f$ & $0.173$ & $0.174$ & $0.191$ & $0.117$ & $\underline{0.224}$ & $0.270$ & $0.224$ & $0.223$ & $0.241$ & $0.139$ & $0.281$ & $\underline{0.375}$ \\
only $\widetilde{\mathbf{M}}_f$ & $0.172$ & $0.172$ & $0.189$ & $0.116$ & $0.216$ & $0.261$ & $0.222$ & $0.222$ & $0.235$ & $0.142$ & $0.284$ & $0.373$ \\
\hline
\ourname & $\mathbf{0.179}$ & $\mathbf{0.178}$ & $\mathbf{0.196}$ & $\mathbf{0.124}$ & $\mathbf{0.227}$ & $\mathbf{0.283}$ & $\mathbf{0.227}$ & $\mathbf{0.227}$ & $\mathbf{0.254}$ & $\mathbf{0.150}$ & $\mathbf{0.298}$ & $\mathbf{0.379}$ \\
\hline
\end{tabular}
    \label{tab:ablation}
\end{table*}

\subsection{Expermental Setting}

\subsection{Datasets}
To verify the effectiveness of the proposed \ourname, we conduct experiments on two public dataset from Douban\footnote{\url{https://www.douban.com/}} and Amazon\footnote{\url{https://jmcauley.ucsd.edu/data/amazon/}}. We select three CDSR scenarios for experiments: ``Book-Movie'' and ``Book-Music'' for Douban and ``Book-Movie'' for Amazon. To ensure sequential constraints, we retain cross-domain interaction sequences that include at least 3 items from each domain within one year. Ratings greater than 3 are considered positive feedback, while ratings less than or equal to 3 are considered negative feedback. We preprocess the data into a sequential recommendation format and split it based on timestamp records: the first 80\% serves as the training set, the next 10\% as the validation set, and the final 10\% as the test set. Statistics are provided in Table~\ref{tab:stat} and Table~\ref{tab:stat2}.

\subsection{Baselines}
In this section, We compare \ourname~ with three representative sequential recommendation (SR) baselines and four cross-domain SR baselines:

\noindent \textit{Sequential recommendation baselines}:
\begin{itemize}[leftmargin=*]
    \item \textbf{GRU4Rec~\cite{hidasi2015session}} uses GRUs to capture sequential patterns in session-based data for personalized recommendations.
    \item \textbf{SASRec~\cite{kang2018self}} uses self-attention to capture long-range dependencies in user behavior for effective recommendations.
    \item \textbf{SRGNN~\cite{wu2019session}} applies Graph Neural Networks (GNN) to session-based recommendation, representing user behaviors as graphs to capture complex relationships and improve accuracy.
\end{itemize}
\textit{Cross-domain sequential recommendation baselines}:
\begin{itemize}[leftmargin=*]
    \item \textbf{CoNet~\cite{hu2018conet}} models interactions in two domains using base networks and transitions information via a cross-network.
    \item \textbf{$\pi$-Net~\cite{ma2019pi}} introducing a novel gating recurrent module to model and transfer knowledge across different domains.
    \item \textbf{C2DSR~\cite{cao2022contrastive}} uses a GNN to leverage inter-domain co-occurrences and employs a contrastive infomax objective to transfer cross-domain preferences by maximizing mutual information.
    \item \textbf{TriCDR~\cite{ma2024triple}} uses triple cross-domain attention and contrastive learning to model comprehensive cross-domain correlations.
    \item \textbf{CA-CDSR~\cite{yin2024learning}} focus on item representation alignment via sequence-aware augmentation and adaptive partial alignment.
\end{itemize}

\subsection{Evaluation Metrics}

To ensure unbiased evaluation, we employ the leave-one-out method, consistent with methodologies used in previous studies~\cite{ye2023dream,kang2018self,zhang2025test,shi2025unified}. Following Rendle's approach~\cite{krichene2020sampled}, each validation/test case is assessed with $1,000$ scores, comprising $999$ negative items and $1$ positive item. The Top-K recommendation performance across these $1,000$ ranking lists is evaluated using metrics such as MRR@$10$ (Mean Reciprocal Rank)~\cite{voorhees1999trec}, NDCG@$5$, $10$ (Normalized Discounted Cumulative Gain)~\cite{jarvelin2002cumulated}, and HR@$1$, $5$, $10$ (Hit Ratio)~\cite{koren2009matrix}, which are common evaluation metrics in recommendation~\cite{zhang2024qagcf, chen2023controllable, wang2024not,shi2024unisar,zhang2024model}.

\subsection{Implementation Details}

Our algorithm is implemented in PyTorch~\cite{paszke2019pytorch}
. The embedding size ($D$) and mini-batch size ($N$) are fixed at $256$, with training epochs set to $100$ and dropout fixed at $0.2$. We employ Adam~\cite{kingma2014adam} as the optimizer for parameter updates. The \(L_2\) regularizer coefficient is chosen from \{$1e\text{-}4$, $5e\text{-}5$, $1e\text{-}5$\}, and the learning rate \(lr\) is selected from \{$1e\text{-}3$, $5e\text{-}4$, $1e\text{-}4$\}. In C2DSR, we vary the depth of the GNN \(L\) from \{$1$, $2$, $3$\}, and adjust the harmonic factor \(\lambda\) from $0.1$ to $0.9$ in increments of $0.1$. For SASRec-based models, we incorporate two single-head attention blocks and learnable position embeddings. The channel number is set to $5$ for \(\pi\)-net, consistent with the original paper~\cite{ma2019pi}. The magnitude of feature augmentation is fixed to $0.1$ in CA-CDSR. \(N_H\) for Head Number is chosen from \{$4$, $8$, $12$, $16$\}.  Best-performing models are determined based on the highest Mean Reciprocal Rank (MRR) performance on the validation set, and their results are reported on the test set. The experimental setup includes Ubuntu 18.04 with an Intel Xeon Gold 5218 CPU (64 cores, 128 threads), 754GB RAM, and four NVIDIA GeForce RTX 3090 GPUs, each with 24GB of memory. The GPU driver version is 550.78, and the CUDA version is 12.4. 

\subsection{Overall Performance}

Table~\ref{tab:exp1}, Table~\ref{tab:exp2} and Table~\ref{tab:exp3} demonstrate the performance of the methods in the ``Book-movie'' and ``Book-music'' of Douban, and ``Book-Movie'' of Amazon, both CDSR scenarios. 

(1) For SR baselines, GRU4Rec, SASRec and SRGNN perform well, with SRGNN performing the best. This validates that modeling interaction sequences using graph neural networks can provide valuable insights for making accurate recommendations. 

(2) For CDSR baselines, most baselines perform better than SR baselines because considering knowledge transfer between domains helps improve cross-domain recommendations. CoNet's performance is slightly lower than SR baselines, possibly due to the weaker capability of its model architecture. Moreover, TriCDR and CA-CDSR based on contrastive learning, exhibit superior performance. This indicates that contrastive learning holds great potential in enhancing representation learning and transferring knowledge within CDSR. 

(3) Our \ourname~ significantly outperforms all baselines across all metrics, demonstrating the superiority of our model in the CDSR task. These experimental results prove that considering domain transition and feedback transition in modeling single-domain and cross-domain recommendations is highly effective.

\begin{figure}[t]
\centering
\includegraphics[width=0.5\textwidth]{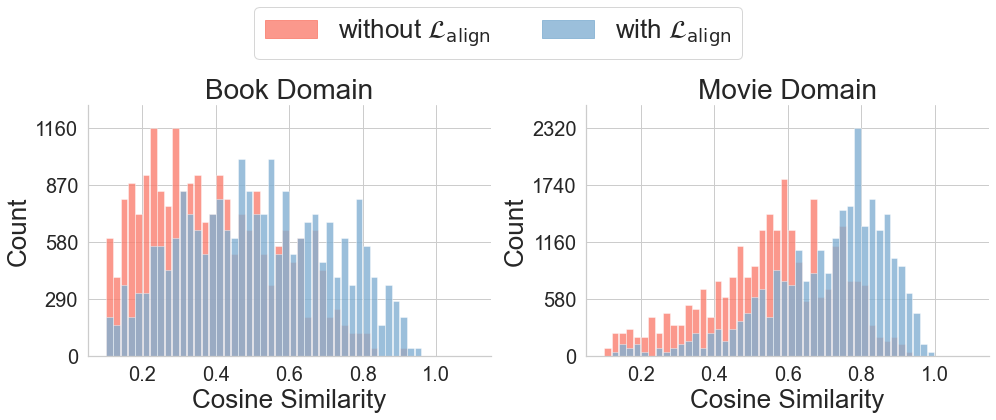}
\caption{Analysis of domain alignment loss $\mathcal{L}_{\text{align}}$.}
\label{ana6}
\end{figure}

\begin{figure}[t]
\centering
\includegraphics[width=0.5\textwidth]{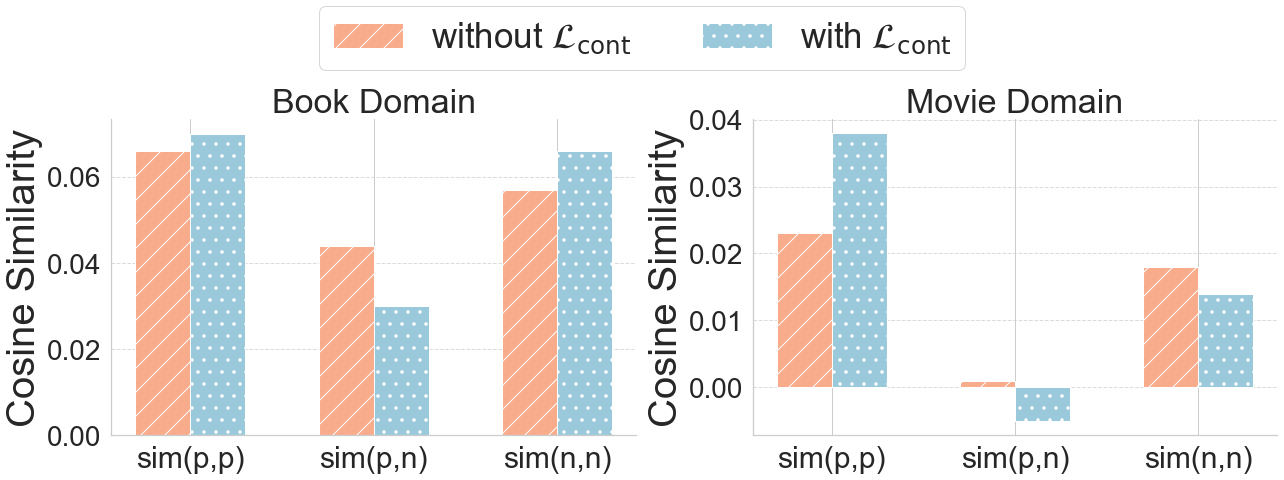}
\caption{Analysis of feedback contrast loss $\mathcal{L}_{\text{cont}}$.}
\label{ana7}
\end{figure}

\subsection{Ablation Studies}


The results of the ablation study, presented in Table~\ref{tab:ablation} for the ``Book-Movie'' scenario, reveal several key insights into the contributions of various components within \ourname:

\begin{itemize}[leftmargin=*]
    \item \textbf{Impact of Transition-Aware Graph Encoder and Cross-Transition MHSA:} The removal of the Transition-Aware Graph Encoder (\textbf{w/o TAG}) and the replacement of Cross-Transition Multi-Head Self-Attention with regular MHSA (\textbf{w/o CTM}) result in a notable decline in recommendation performance. This clearly demonstrates the critical role these components play in capturing the transitions between different domains. Their combined absence highlights their synergistic importance in maintaining high performance.

    \item \textbf{Impact of Transition Alignment and Contrast:} Removing intra-domain embedding alignment (\textbf{w/o $\mathcal{L}_{\text{align}}$}) or in-sequence positive-negative sample contrast (\textbf{w/o $\mathcal{L}_{\text{cont}}$}) leads to a significant decline in recommendation performance. This clearly demonstrates that embeddings within the same domain should exhibit stronger similarity, and positive/negative items from each user interaction should be differentiated to a greater extent. Both mechanisms contribute to improving the model's performance by enhancing representation quality and discriminative power.

    \item \textbf{Effect of Retaining Individual Transition Masks in Cross-Domain Sequence:} When the model is configured to retain only individual transition masks in the MHSA of the cross-domain sequence (\textbf{only $\mathbf{M}_1$}, \textbf{only $\mathbf{M}_2$}, \textbf{only $\mathbf{M}_3$}, \textbf{only $\mathbf{M}_4$}), the general trend is a degradation in performance. This indicates that each transition mask contributes distinct information, and isolating them misses important cross-domain transitions. However, an interesting exception is observed with \textbf{only $\mathbf{M}_1$} in the Movie domain, where a slight improvement than others is noted. This could imply that in certain cases, specific transitions might hold more relevance, but the overall performance gain from retaining all masks indicates that the comprehensive transition information is essential for robust recommendations.

    \item \textbf{Performance with Single-Domain Sequence Masks:} In the MHSA of the single-domain sequence, retaining only feedback transition information (\textbf{only $\mathbf{M}_f$}) or only non-feedback transition information (\textbf{only $\widetilde{\mathbf{M}}_f$}) leads to a reduction in performance. This finding underscores the importance of the mask design in \ourname, which balances both transition and non-transition information. The decline in performance when these elements are isolated suggests that both types of information are integral to accurately modeling user preferences within a single domain, further validating the effectiveness of the dual-mask approach in \ourname.
\end{itemize}

\subsection{\mbox{Analysis of Transition Alignment and Contrast}}
\subsubsection{Analysis of Domain Alignment Loss $\mathcal{L}_{\text{align}}$}
To investigate the impact of the domain alignment loss $\mathcal{L}_{\text{align}}$ described in Section~\ref{cl} on cross-domain representations, we first analyze its alignment effectiveness. As shown in Figure~\ref{ana6}, we compare the cosine similarity distributions between Book domain embeddings $\mathbf{E}_{A}$ and Cross-domain embeddings $\mathbf{E}_{C}$, as well as between Movie domain embeddings $\mathbf{E}_{B}$ and $\mathbf{E}_{C}$, under conditions with and without $\mathcal{L}_{\text{align}}$. The significantly lower similarity observed without $\mathcal{L}_{\text{align}}$ demonstrates that our alignment loss effectively reduces embedding discrepancies caused by different encoders, thereby ensuring consistent cross-domain representation learning.

\subsubsection{Analysis of Feedback Contrast Loss $\mathcal{L}_{\text{cont}}$}
We further examine the effect of feedback contrast loss $\mathcal{L}_{\text{cont}}$ through inter-item similarity analysis. As illustrated in Figure~\ref{ana7}, we calculate three similarity metrics using embeddings from Book domain: positive-positive similarity ($sim(p,p)$), positive-negative similarity ($sim(p,n)$), and negative-negative similarity ($sim(n,n)$). The results reveal that with $\mathcal{L}_{\text{cont}}$, same-feedback items exhibit enhanced intra-cluster similarity while showing reduced inter-cluster similarity (even reaching negative values in Movie domain). This contrastive pattern confirms that $\mathcal{L}_{\text{cont}}$ successfully differentiates user preference signals in interaction sequences, leading to more discriminative representations for recommendation improvement.

\subsection{Sensitive Analysis of Hyper-parameters}
\begin{figure}[t]
\centering
\includegraphics[width=0.5\textwidth]{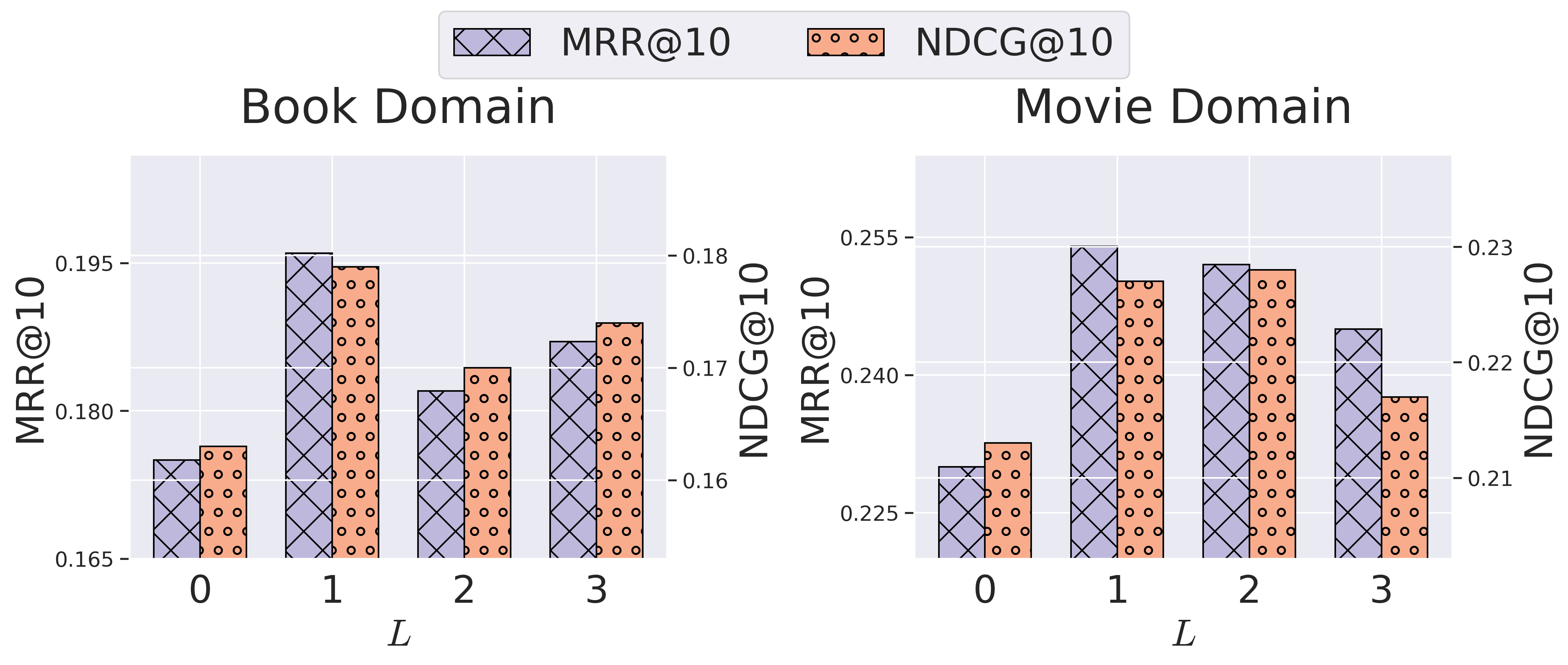}
\caption{
Sensitivity Analysis of $L$.}
\label{ana1}
\end{figure}

\begin{figure}[t]
\centering
\includegraphics[width=0.5\textwidth]{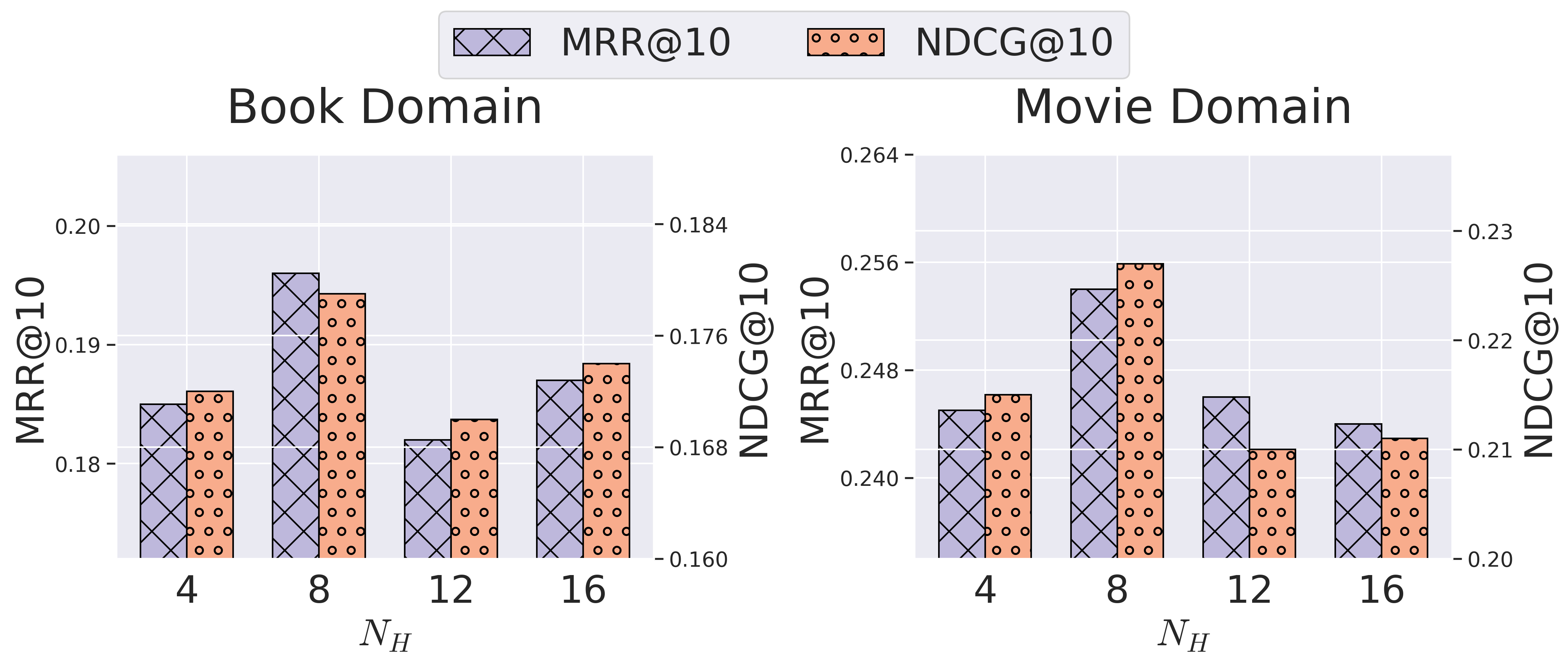}
\caption{
Sensitivity Analysis of $N_H$.}
\label{ana2}
\end{figure}

In this section, we conducted more detailed sensitive analysis of the hyper-parameters setting in \ourname.
\subsubsection{Analysis of \(L\) and \(N_H\)}
This section investigates the parameter sensitivity of the depth \(L\) of the graph encoder and the number of self-attention heads \(N_H\). For fair comparison, when studying \(L\), we fix \(N_H = 8\); when studying \(N_H\), we fix \(L = 1\). For the hyperparameter \(L\), we conducted analysis experiments with values in \{0, 1, 2, 3\}, where $L = 0$ means ignoring the Transition-Aware Graph Encoder. Figure~\ref{ana1} shows the impact on recommendation performance in terms of MRR@10 and NDCG@10 metrics for the ``Book" and ``Movie" domains on Douban, where it is observed that our model achieves the best performance when \(L = 1\) on Movie domain. For the hyperparameter \(N_H\), we selected values from \{4, 8, 12, 16\}, and Figure~\ref{ana2} show the impact on recommendation performance in terms of MRR@10 and NDCG@10 metrics for the "Book" and "Movie" domains, where \ourname~ achieves the best recommendation performance on Book domain when \(N_H = 8\).

\begin{figure}[t]
\centering
\includegraphics[width=0.5\textwidth]{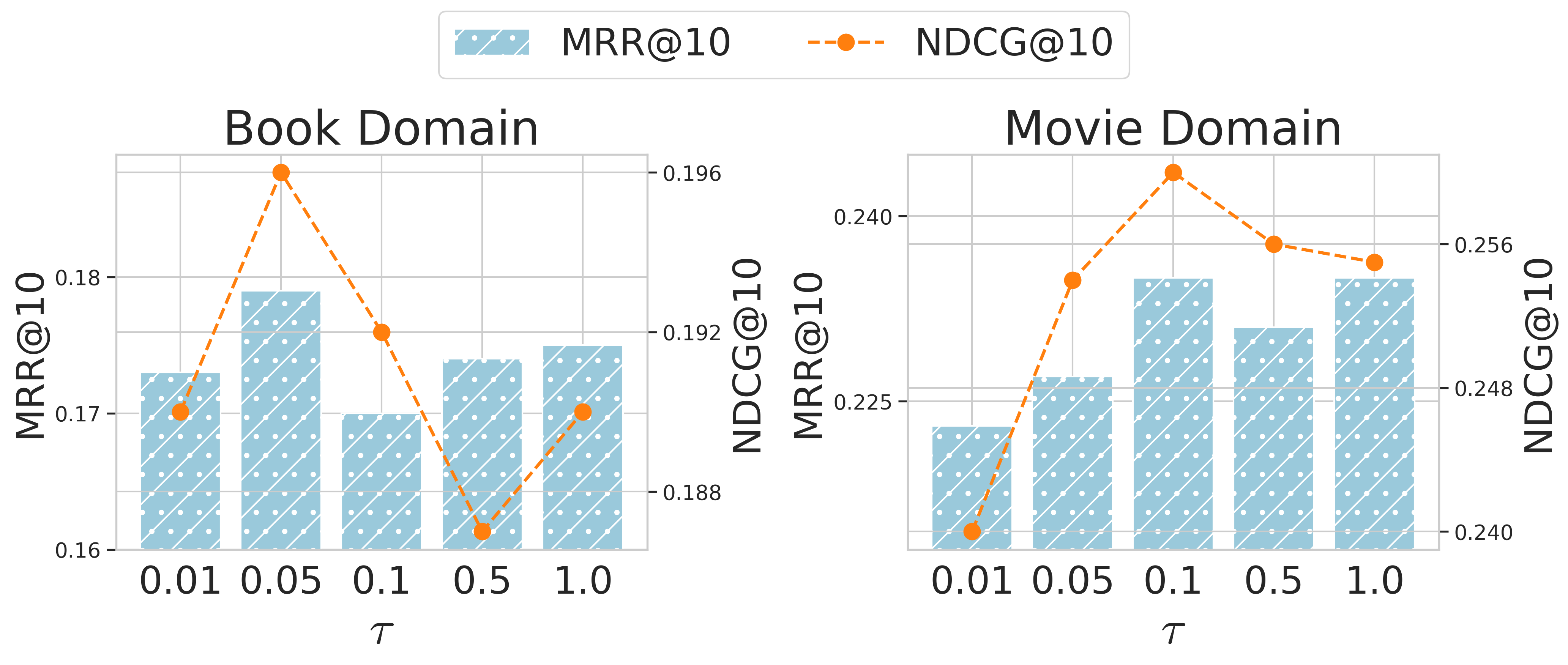}
\caption{
Sensitivity Analysis of $\tau$.}
\label{ana3}
\end{figure}

\begin{figure}[t]
\centering
\includegraphics[width=0.5\textwidth]{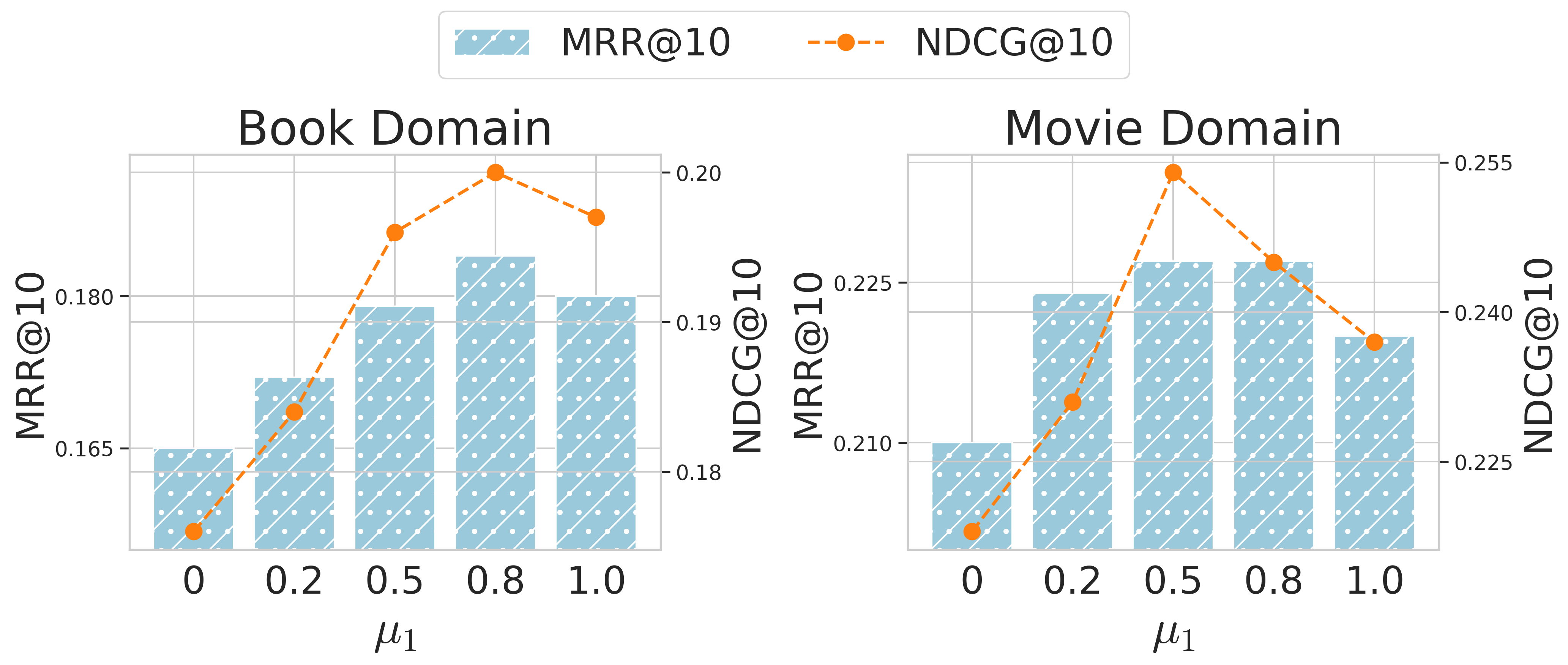}
\caption{
Sensitivity Analysis of $\mu_1$.}
\label{ana4}
\end{figure}

\begin{figure}[t]
\centering
\includegraphics[width=0.5\textwidth]{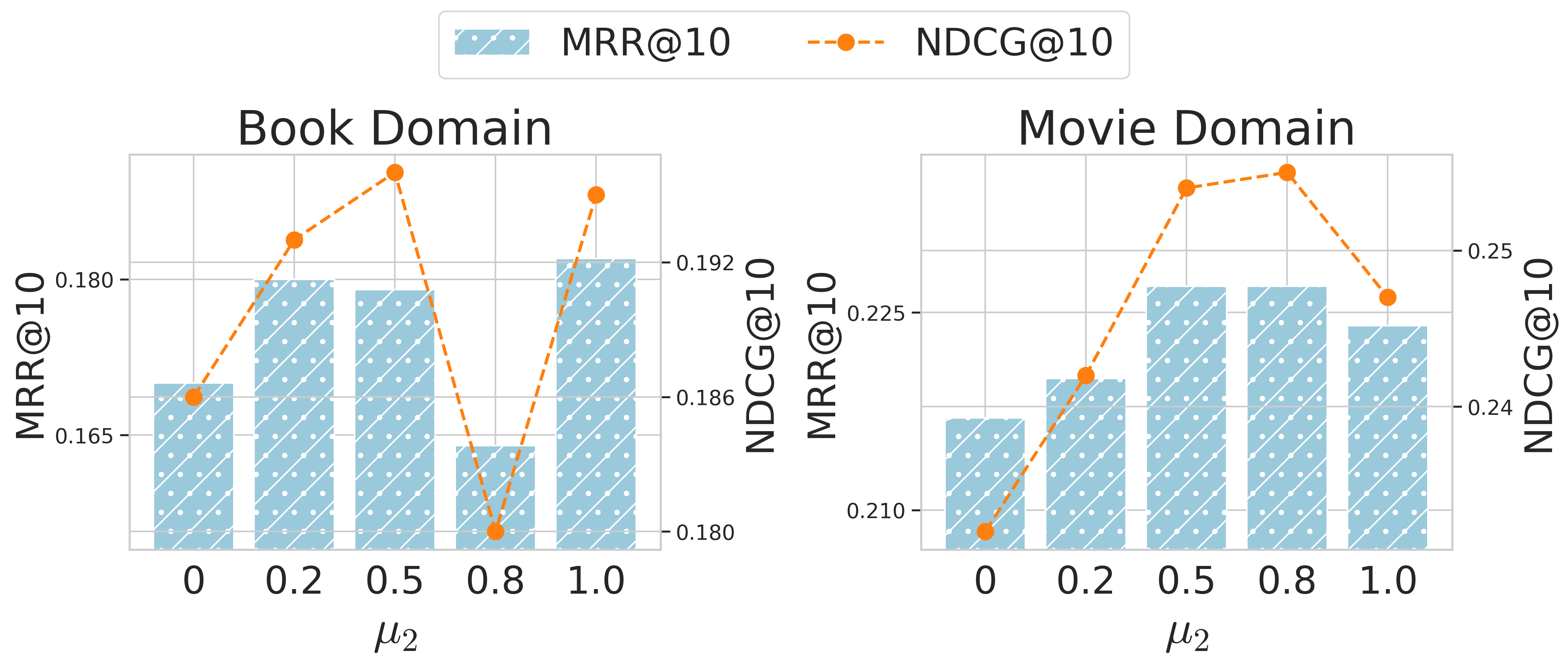}
\caption{
Sensitivity Analysis of $\mu_2$.}
\label{ana5}
\end{figure}

\subsubsection{Analysis of $\tau$, $\mu_1$ and $
\mu_2$}
We analyze the impact of parameters $\tau$, $\mu_1$ and $\mu_2$ in the contrastive learning loss $\mathcal{L}_{\text{align}}$ and $\mathcal{L}_{\text{cont}}$. For fair comparison, when studying $\mu_1$ and $\mu_2$ with values in $\{0, 0.2, 0.5, 0.7, 1.0\}$, we fix $\tau=0.1$. $\mu_1=0$ and $\mu_2=0$ mean ignore the domain alignment loss $\mathcal{L}_{\text{align}}$ and the feedback contrast loss $\mathcal{L}_{\text{cont}}$, respectively; when studying $\tau$ with values in $\{0.01, 0.05, 0.1, 0.5, 1.0\}$, we fix $\mu_1=0.5$ and $\mu_2=0.5$. Figure~\ref{ana3}, Figure~\ref{ana4}, Figure~\ref{ana5} shows the impact of $\tau$, $\mu_1$, $\mu_2$ on recommendation performance in terms of MRR@10 and NDCG@10 metrics for the ``Book" and ``Movie" domains on Douban, respectively, where it can be observed from the figures that the optimal parameter settings on Book-Movie of Douban are observed to be $\mu_1=0.5$, $\mu_2=0.5$ and $\tau=0.1$. The results demonstrate that \ourname~is quite sensitive to the hyper-parameters, and both too large and too small values of these parameters lead to a decrease in recommendation performance.

%% file: section/5.conclusion.tex
\section{Conclusion}
Our work addresses the limitations of current recommendation systems that predominantly focus on domain transitions by proposing \ourname, a novel method to model user interest transitions across different domains and feedback types. By introducing a transition-aware graph encoder that assigns varying weights to edges based on feedback type, \ourname~ effectively captures the nuanced transition information in user history. The use of cross-transition multi-head self-attention further refines these embeddings by distinguishing different types of transitions with various masks. To further enhance representation learning, two contrastive losses are utilized to align transitions across domains and feedback types. Our experimental results on two public datasets validate the effectiveness of \ourname, demonstrating its capability to enhance prediction accuracy across multiple domains. This approach not only advances the modeling of user interests in recommendation systems but also highlights the importance of incorporating both positive and negative feedback in cross-domain transitions.



%% file: main.bbl

\begin{thebibliography}{43}


\ifx \showCODEN    \undefined \def \showCODEN     #1{\unskip}     \fi
\ifx \showISBNx    \undefined \def \showISBNx     #1{\unskip}     \fi
\ifx \showISBNxiii \undefined \def \showISBNxiii  #1{\unskip}     \fi
\ifx \showISSN     \undefined \def \showISSN      #1{\unskip}     \fi
\ifx \showLCCN     \undefined \def \showLCCN      #1{\unskip}     \fi
\ifx \shownote     \undefined \def \shownote      #1{#1}          \fi
\ifx \showarticletitle \undefined \def \showarticletitle #1{#1}   \fi
\ifx \showURL      \undefined \def \showURL       {\relax}        \fi
\providecommand\bibfield[2]{#2}
\providecommand\bibinfo[2]{#2}
\providecommand\natexlab[1]{#1}
\providecommand\showeprint[2][]{arXiv:#2}

\bibitem[Anwar et~al\mbox{.}(2019)]%
        {anwar2019machine}
\bibfield{author}{\bibinfo{person}{Khalid Anwar}, \bibinfo{person}{Jamshed Siddiqui}, {and} \bibinfo{person}{Shahab Saquib~Sohail}.} \bibinfo{year}{2019}\natexlab{}.
\newblock \showarticletitle{Machine learning techniques for book recommendation: an overview}. In \bibinfo{booktitle}{\emph{Proceedings of International Conference on Sustainable Computing in Science, Technology and Management (SUSCOM), Amity University Rajasthan, Jaipur-India}}.
\newblock


\bibitem[Cao et~al\mbox{.}(2022)]%
        {cao2022contrastive}
\bibfield{author}{\bibinfo{person}{Jiangxia Cao}, \bibinfo{person}{Xin Cong}, \bibinfo{person}{Jiawei Sheng}, \bibinfo{person}{Tingwen Liu}, {and} \bibinfo{person}{Bin Wang}.} \bibinfo{year}{2022}\natexlab{}.
\newblock \showarticletitle{Contrastive cross-domain sequential recommendation}. In \bibinfo{booktitle}{\emph{Proceedings of the 31st ACM International Conference on Information \& Knowledge Management}}. \bibinfo{pages}{138--147}.
\newblock


\bibitem[Chen et~al\mbox{.}(2019)]%
        {chen2019dagcn}
\bibfield{author}{\bibinfo{person}{Fengwen Chen}, \bibinfo{person}{Shirui Pan}, \bibinfo{person}{Jing Jiang}, \bibinfo{person}{Huan Huo}, {and} \bibinfo{person}{Guodong Long}.} \bibinfo{year}{2019}\natexlab{}.
\newblock \showarticletitle{DAGCN: dual attention graph convolutional networks}. In \bibinfo{booktitle}{\emph{2019 International Joint Conference on Neural Networks (IJCNN)}}. IEEE, \bibinfo{pages}{1--8}.
\newblock


\bibitem[Chen et~al\mbox{.}(2023)]%
        {chen2023controllable}
\bibfield{author}{\bibinfo{person}{Sirui Chen}, \bibinfo{person}{Yuan Wang}, \bibinfo{person}{Zijing Wen}, \bibinfo{person}{Zhiyu Li}, \bibinfo{person}{Changshuo Zhang}, \bibinfo{person}{Xiao Zhang}, \bibinfo{person}{Quan Lin}, \bibinfo{person}{Cheng Zhu}, {and} \bibinfo{person}{Jun Xu}.} \bibinfo{year}{2023}\natexlab{}.
\newblock \showarticletitle{Controllable multi-objective re-ranking with policy hypernetworks}. In \bibinfo{booktitle}{\emph{Proceedings of the 29th ACM SIGKDD conference on knowledge discovery and data mining}}. \bibinfo{pages}{3855--3864}.
\newblock


\bibitem[Goyani and Chaurasiya(2020)]%
        {goyani2020review}
\bibfield{author}{\bibinfo{person}{Mahesh Goyani} {and} \bibinfo{person}{Neha Chaurasiya}.} \bibinfo{year}{2020}\natexlab{}.
\newblock \showarticletitle{A review of movie recommendation system: Limitations, Survey and Challenges}.
\newblock \bibinfo{journal}{\emph{ELCVIA: electronic letters on computer vision and image analysis}} \bibinfo{volume}{19}, \bibinfo{number}{3} (\bibinfo{year}{2020}), \bibinfo{pages}{0018--37}.
\newblock


\bibitem[Hidasi et~al\mbox{.}(2015)]%
        {hidasi2015session}
\bibfield{author}{\bibinfo{person}{Bal{\'a}zs Hidasi}, \bibinfo{person}{Alexandros Karatzoglou}, \bibinfo{person}{Linas Baltrunas}, {and} \bibinfo{person}{Domonkos Tikk}.} \bibinfo{year}{2015}\natexlab{}.
\newblock \showarticletitle{Session-based recommendations with recurrent neural networks}.
\newblock \bibinfo{journal}{\emph{arXiv preprint arXiv:1511.06939}} (\bibinfo{year}{2015}).
\newblock


\bibitem[Hu et~al\mbox{.}(2018)]%
        {hu2018conet}
\bibfield{author}{\bibinfo{person}{Guangneng Hu}, \bibinfo{person}{Yu Zhang}, {and} \bibinfo{person}{Qiang Yang}.} \bibinfo{year}{2018}\natexlab{}.
\newblock \showarticletitle{Conet: Collaborative cross networks for cross-domain recommendation}. In \bibinfo{booktitle}{\emph{Proceedings of the 27th ACM international conference on information and knowledge management}}. \bibinfo{pages}{667--676}.
\newblock


\bibitem[J{\"a}rvelin and Kek{\"a}l{\"a}inen(2002)]%
        {jarvelin2002cumulated}
\bibfield{author}{\bibinfo{person}{Kalervo J{\"a}rvelin} {and} \bibinfo{person}{Jaana Kek{\"a}l{\"a}inen}.} \bibinfo{year}{2002}\natexlab{}.
\newblock \showarticletitle{Cumulated gain-based evaluation of IR techniques}.
\newblock \bibinfo{journal}{\emph{ACM Transactions on Information Systems (TOIS)}} \bibinfo{volume}{20}, \bibinfo{number}{4} (\bibinfo{year}{2002}), \bibinfo{pages}{422--446}.
\newblock


\bibitem[Kang and McAuley(2018)]%
        {kang2018self}
\bibfield{author}{\bibinfo{person}{Wang-Cheng Kang} {and} \bibinfo{person}{Julian McAuley}.} \bibinfo{year}{2018}\natexlab{}.
\newblock \showarticletitle{Self-attentive sequential recommendation}. In \bibinfo{booktitle}{\emph{2018 IEEE international conference on data mining (ICDM)}}. IEEE, \bibinfo{pages}{197--206}.
\newblock


\bibitem[Kingma and Ba(2014)]%
        {kingma2014adam}
\bibfield{author}{\bibinfo{person}{Diederik~P Kingma} {and} \bibinfo{person}{Jimmy Ba}.} \bibinfo{year}{2014}\natexlab{}.
\newblock \showarticletitle{Adam: A method for stochastic optimization}.
\newblock \bibinfo{journal}{\emph{arXiv preprint arXiv:1412.6980}} (\bibinfo{year}{2014}).
\newblock


\bibitem[Koren et~al\mbox{.}(2009)]%
        {koren2009matrix}
\bibfield{author}{\bibinfo{person}{Yehuda Koren}, \bibinfo{person}{Robert Bell}, {and} \bibinfo{person}{Chris Volinsky}.} \bibinfo{year}{2009}\natexlab{}.
\newblock \showarticletitle{Matrix factorization techniques for recommender systems}.
\newblock \bibinfo{journal}{\emph{Computer}} \bibinfo{volume}{42}, \bibinfo{number}{8} (\bibinfo{year}{2009}), \bibinfo{pages}{30--37}.
\newblock


\bibitem[Krichene and Rendle(2020)]%
        {krichene2020sampled}
\bibfield{author}{\bibinfo{person}{Walid Krichene} {and} \bibinfo{person}{Steffen Rendle}.} \bibinfo{year}{2020}\natexlab{}.
\newblock \showarticletitle{On sampled metrics for item recommendation}. In \bibinfo{booktitle}{\emph{Proceedings of the 26th ACM SIGKDD international conference on knowledge discovery \& data mining}}. \bibinfo{pages}{1748--1757}.
\newblock


\bibitem[Lei et~al\mbox{.}(2021)]%
        {lei2021semi}
\bibfield{author}{\bibinfo{person}{Chenyi Lei}, \bibinfo{person}{Yong Liu}, \bibinfo{person}{Lingzi Zhang}, \bibinfo{person}{Guoxin Wang}, \bibinfo{person}{Haihong Tang}, \bibinfo{person}{Houqiang Li}, {and} \bibinfo{person}{Chunyan Miao}.} \bibinfo{year}{2021}\natexlab{}.
\newblock \showarticletitle{Semi: A sequential multi-modal information transfer network for e-commerce micro-video recommendations}. In \bibinfo{booktitle}{\emph{Proceedings of the 27th ACM SIGKDD Conference on Knowledge Discovery \& Data Mining}}. \bibinfo{pages}{3161--3171}.
\newblock


\bibitem[Li et~al\mbox{.}(2022)]%
        {li2022recguru}
\bibfield{author}{\bibinfo{person}{Chenglin Li}, \bibinfo{person}{Mingjun Zhao}, \bibinfo{person}{Huanming Zhang}, \bibinfo{person}{Chenyun Yu}, \bibinfo{person}{Lei Cheng}, \bibinfo{person}{Guoqiang Shu}, \bibinfo{person}{Beibei Kong}, {and} \bibinfo{person}{Di Niu}.} \bibinfo{year}{2022}\natexlab{}.
\newblock \showarticletitle{RecGURU: Adversarial learning of generalized user representations for cross-domain recommendation}. In \bibinfo{booktitle}{\emph{Proceedings of the fifteenth ACM international conference on web search and data mining}}. \bibinfo{pages}{571--581}.
\newblock


\bibitem[Li et~al\mbox{.}(2021)]%
        {li2021dual}
\bibfield{author}{\bibinfo{person}{Pan Li}, \bibinfo{person}{Zhichao Jiang}, \bibinfo{person}{Maofei Que}, \bibinfo{person}{Yao Hu}, {and} \bibinfo{person}{Alexander Tuzhilin}.} \bibinfo{year}{2021}\natexlab{}.
\newblock \showarticletitle{Dual attentive sequential learning for cross-domain click-through rate prediction}. In \bibinfo{booktitle}{\emph{Proceedings of the 27th ACM SIGKDD conference on knowledge discovery \& data mining}}. \bibinfo{pages}{3172--3180}.
\newblock


\bibitem[Ma et~al\mbox{.}(2024)]%
        {ma2024triple}
\bibfield{author}{\bibinfo{person}{Haokai Ma}, \bibinfo{person}{Ruobing Xie}, \bibinfo{person}{Lei Meng}, \bibinfo{person}{Xin Chen}, \bibinfo{person}{Xu Zhang}, \bibinfo{person}{Leyu Lin}, {and} \bibinfo{person}{Jie Zhou}.} \bibinfo{year}{2024}\natexlab{}.
\newblock \showarticletitle{Triple sequence learning for cross-domain recommendation}.
\newblock \bibinfo{journal}{\emph{ACM Transactions on Information Systems}} \bibinfo{volume}{42}, \bibinfo{number}{4} (\bibinfo{year}{2024}), \bibinfo{pages}{1--29}.
\newblock


\bibitem[Ma et~al\mbox{.}(2022)]%
        {ma2022mixed}
\bibfield{author}{\bibinfo{person}{Muyang Ma}, \bibinfo{person}{Pengjie Ren}, \bibinfo{person}{Zhumin Chen}, \bibinfo{person}{Zhaochun Ren}, \bibinfo{person}{Lifan Zhao}, \bibinfo{person}{Peiyu Liu}, \bibinfo{person}{Jun Ma}, {and} \bibinfo{person}{Maarten de Rijke}.} \bibinfo{year}{2022}\natexlab{}.
\newblock \showarticletitle{Mixed information flow for cross-domain sequential recommendations}.
\newblock \bibinfo{journal}{\emph{ACM Transactions on Knowledge Discovery from Data (TKDD)}} \bibinfo{volume}{16}, \bibinfo{number}{4} (\bibinfo{year}{2022}), \bibinfo{pages}{1--32}.
\newblock


\bibitem[Ma et~al\mbox{.}(2019)]%
        {ma2019pi}
\bibfield{author}{\bibinfo{person}{Muyang Ma}, \bibinfo{person}{Pengjie Ren}, \bibinfo{person}{Yujie Lin}, \bibinfo{person}{Zhumin Chen}, \bibinfo{person}{Jun Ma}, {and} \bibinfo{person}{Maarten~de Rijke}.} \bibinfo{year}{2019}\natexlab{}.
\newblock \showarticletitle{$\pi$-net: A parallel information-sharing network for shared-account cross-domain sequential recommendations}. In \bibinfo{booktitle}{\emph{Proceedings of the 42nd international ACM SIGIR conference on research and development in information retrieval}}. \bibinfo{pages}{685--694}.
\newblock


\bibitem[Oord et~al\mbox{.}(2018)]%
        {oord2018representation}
\bibfield{author}{\bibinfo{person}{Aaron van~den Oord}, \bibinfo{person}{Yazhe Li}, {and} \bibinfo{person}{Oriol Vinyals}.} \bibinfo{year}{2018}\natexlab{}.
\newblock \showarticletitle{Representation learning with contrastive predictive coding}.
\newblock \bibinfo{journal}{\emph{arXiv preprint arXiv:1807.03748}} (\bibinfo{year}{2018}).
\newblock


\bibitem[Paszke et~al\mbox{.}(2019)]%
        {paszke2019pytorch}
\bibfield{author}{\bibinfo{person}{Adam Paszke}, \bibinfo{person}{Sam Gross}, \bibinfo{person}{Francisco Massa}, \bibinfo{person}{Adam Lerer}, \bibinfo{person}{James Bradbury}, \bibinfo{person}{Gregory Chanan}, \bibinfo{person}{Trevor Killeen}, \bibinfo{person}{Zeming Lin}, \bibinfo{person}{Natalia Gimelshein}, \bibinfo{person}{Luca Antiga}, {et~al\mbox{.}}} \bibinfo{year}{2019}\natexlab{}.
\newblock \showarticletitle{Pytorch: An imperative style, high-performance deep learning library}.
\newblock \bibinfo{journal}{\emph{Advances in neural information processing systems}}  \bibinfo{volume}{32} (\bibinfo{year}{2019}).
\newblock


\bibitem[Perera and Zimmermann(2020a)]%
        {perera2020lstm}
\bibfield{author}{\bibinfo{person}{Dilruk Perera} {and} \bibinfo{person}{Roger Zimmermann}.} \bibinfo{year}{2020}\natexlab{a}.
\newblock \showarticletitle{LSTM networks for online cross-network recommendations}.
\newblock \bibinfo{journal}{\emph{arXiv preprint arXiv:2008.10849}} (\bibinfo{year}{2020}).
\newblock


\bibitem[Perera and Zimmermann(2020b)]%
        {perera2020towards}
\bibfield{author}{\bibinfo{person}{Dilruk Perera} {and} \bibinfo{person}{Roger Zimmermann}.} \bibinfo{year}{2020}\natexlab{b}.
\newblock \showarticletitle{Towards comprehensive recommender systems: Time-aware unified recommendations based on listwise ranking of implicit cross-network data}. In \bibinfo{booktitle}{\emph{Proceedings of the AAAI Conference on Artificial Intelligence}}, Vol.~\bibinfo{volume}{34}. \bibinfo{pages}{189--197}.
\newblock


\bibitem[Qu et~al\mbox{.}(2024)]%
        {qu2024budgeted}
\bibfield{author}{\bibinfo{person}{Yunke Qu}, \bibinfo{person}{Tong Chen}, \bibinfo{person}{Quoc Viet~Hung Nguyen}, {and} \bibinfo{person}{Hongzhi Yin}.} \bibinfo{year}{2024}\natexlab{}.
\newblock \showarticletitle{Budgeted embedding table for recommender systems}. In \bibinfo{booktitle}{\emph{Proceedings of the 17th ACM International Conference on Web Search and Data Mining}}. \bibinfo{pages}{557--566}.
\newblock


\bibitem[Qu et~al\mbox{.}(2023)]%
        {qu2023continuous}
\bibfield{author}{\bibinfo{person}{Yunke Qu}, \bibinfo{person}{Tong Chen}, \bibinfo{person}{Xiangyu Zhao}, \bibinfo{person}{Lizhen Cui}, \bibinfo{person}{Kai Zheng}, {and} \bibinfo{person}{Hongzhi Yin}.} \bibinfo{year}{2023}\natexlab{}.
\newblock \showarticletitle{Continuous input embedding size search for recommender systems}. In \bibinfo{booktitle}{\emph{Proceedings of the 46th International ACM SIGIR Conference on Research and Development in Information Retrieval}}. \bibinfo{pages}{708--717}.
\newblock


\bibitem[Shi et~al\mbox{.}(2024)]%
        {shi2024unisar}
\bibfield{author}{\bibinfo{person}{Teng Shi}, \bibinfo{person}{Zihua Si}, \bibinfo{person}{Jun Xu}, \bibinfo{person}{Xiao Zhang}, \bibinfo{person}{Xiaoxue Zang}, \bibinfo{person}{Kai Zheng}, \bibinfo{person}{Dewei Leng}, \bibinfo{person}{Yanan Niu}, {and} \bibinfo{person}{Yang Song}.} \bibinfo{year}{2024}\natexlab{}.
\newblock \showarticletitle{UniSAR: Modeling User Transition Behaviors between Search and Recommendation}. In \bibinfo{booktitle}{\emph{Proceedings of the 47th International ACM SIGIR Conference on Research and Development in Information Retrieval}}. \bibinfo{pages}{1029--1039}.
\newblock


\bibitem[Shi et~al\mbox{.}(2025)]%
        {shi2025unified}
\bibfield{author}{\bibinfo{person}{Teng Shi}, \bibinfo{person}{Jun Xu}, \bibinfo{person}{Xiao Zhang}, \bibinfo{person}{Xiaoxue Zang}, \bibinfo{person}{Kai Zheng}, \bibinfo{person}{Yang Song}, {and} \bibinfo{person}{Enyun Yu}.} \bibinfo{year}{2025}\natexlab{}.
\newblock \showarticletitle{Unified Generative Search and Recommendation}.
\newblock \bibinfo{journal}{\emph{arXiv preprint arXiv:2504.05730}} (\bibinfo{year}{2025}).
\newblock


\bibitem[Tang et~al\mbox{.}(2025)]%
        {tang2025think}
\bibfield{author}{\bibinfo{person}{Jiakai Tang}, \bibinfo{person}{Sunhao Dai}, \bibinfo{person}{Teng Shi}, \bibinfo{person}{Jun Xu}, \bibinfo{person}{Xu Chen}, \bibinfo{person}{Wen Chen}, \bibinfo{person}{Wu Jian}, {and} \bibinfo{person}{Yuning Jiang}.} \bibinfo{year}{2025}\natexlab{}.
\newblock \showarticletitle{Think Before Recommend: Unleashing the Latent Reasoning Power for Sequential Recommendation}.
\newblock \bibinfo{journal}{\emph{arXiv preprint arXiv:2503.22675}} (\bibinfo{year}{2025}).
\newblock


\bibitem[Voorhees et~al\mbox{.}(1999)]%
        {voorhees1999trec}
\bibfield{author}{\bibinfo{person}{Ellen~M Voorhees} {et~al\mbox{.}}} \bibinfo{year}{1999}\natexlab{}.
\newblock \showarticletitle{The trec-8 question answering track report.}. In \bibinfo{booktitle}{\emph{Trec}}, Vol.~\bibinfo{volume}{99}. \bibinfo{pages}{77--82}.
\newblock


\bibitem[Wang et~al\mbox{.}(2024)]%
        {wang2024not}
\bibfield{author}{\bibinfo{person}{Yuan Wang}, \bibinfo{person}{Zhiyu Li}, \bibinfo{person}{Changshuo Zhang}, \bibinfo{person}{Sirui Chen}, \bibinfo{person}{Xiao Zhang}, \bibinfo{person}{Jun Xu}, {and} \bibinfo{person}{Quan Lin}.} \bibinfo{year}{2024}\natexlab{}.
\newblock \showarticletitle{Do Not Wait: Learning Re-Ranking Model Without User Feedback At Serving Time in E-Commerce}. In \bibinfo{booktitle}{\emph{Proceedings of the 18th ACM Conference on Recommender Systems}}. \bibinfo{pages}{896--901}.
\newblock


\bibitem[Wu et~al\mbox{.}(2019)]%
        {wu2019session}
\bibfield{author}{\bibinfo{person}{Shu Wu}, \bibinfo{person}{Yuyuan Tang}, \bibinfo{person}{Yanqiao Zhu}, \bibinfo{person}{Liang Wang}, \bibinfo{person}{Xing Xie}, {and} \bibinfo{person}{Tieniu Tan}.} \bibinfo{year}{2019}\natexlab{}.
\newblock \showarticletitle{Session-based recommendation with graph neural networks}. In \bibinfo{booktitle}{\emph{Proceedings of the AAAI conference on artificial intelligence}}, Vol.~\bibinfo{volume}{33}. \bibinfo{pages}{346--353}.
\newblock


\bibitem[Xu et~al\mbox{.}(2019)]%
        {xu2019graph}
\bibfield{author}{\bibinfo{person}{Chengfeng Xu}, \bibinfo{person}{Pengpeng Zhao}, \bibinfo{person}{Yanchi Liu}, \bibinfo{person}{Victor~S Sheng}, \bibinfo{person}{Jiajie Xu}, \bibinfo{person}{Fuzhen Zhuang}, \bibinfo{person}{Junhua Fang}, {and} \bibinfo{person}{Xiaofang Zhou}.} \bibinfo{year}{2019}\natexlab{}.
\newblock \showarticletitle{Graph contextualized self-attention network for session-based recommendation.}. In \bibinfo{booktitle}{\emph{IJCAI}}, Vol.~\bibinfo{volume}{19}. \bibinfo{pages}{3940--3946}.
\newblock


\bibitem[Ye et~al\mbox{.}(2023)]%
        {ye2023dream}
\bibfield{author}{\bibinfo{person}{Xiaoxin Ye}, \bibinfo{person}{Yun Li}, {and} \bibinfo{person}{Lina Yao}.} \bibinfo{year}{2023}\natexlab{}.
\newblock \showarticletitle{DREAM: Decoupled Representation via Extraction Attention Module and Supervised Contrastive Learning for Cross-Domain Sequential Recommender}. In \bibinfo{booktitle}{\emph{Proceedings of the 17th ACM Conference on Recommender Systems}}. \bibinfo{pages}{479--490}.
\newblock


\bibitem[Yin et~al\mbox{.}(2024)]%
        {yin2024learning}
\bibfield{author}{\bibinfo{person}{Mingjia Yin}, \bibinfo{person}{Hao Wang}, \bibinfo{person}{Wei Guo}, \bibinfo{person}{Yong Liu}, \bibinfo{person}{Zhi Li}, \bibinfo{person}{Sirui Zhao}, \bibinfo{person}{Zhen Wang}, \bibinfo{person}{Defu Lian}, {and} \bibinfo{person}{Enhong Chen}.} \bibinfo{year}{2024}\natexlab{}.
\newblock \showarticletitle{Learning partially aligned item representation for cross-domain sequential recommendation}.
\newblock \bibinfo{journal}{\emph{arXiv preprint arXiv:2405.12473}} (\bibinfo{year}{2024}).
\newblock


\bibitem[Zhang et~al\mbox{.}(2024a)]%
        {zhang2024reinforcing}
\bibfield{author}{\bibinfo{person}{Changshuo Zhang}, \bibinfo{person}{Sirui Chen}, \bibinfo{person}{Xiao Zhang}, \bibinfo{person}{Sunhao Dai}, \bibinfo{person}{Weijie Yu}, {and} \bibinfo{person}{Jun Xu}.} \bibinfo{year}{2024}\natexlab{a}.
\newblock \showarticletitle{Reinforcing Long-Term Performance in Recommender Systems with User-Oriented Exploration Policy}. In \bibinfo{booktitle}{\emph{Proceedings of the 47th International ACM SIGIR Conference on Research and Development in Information Retrieval}}. \bibinfo{pages}{1850--1860}.
\newblock


\bibitem[Zhang et~al\mbox{.}(2025a)]%
        {zhang2025comment}
\bibfield{author}{\bibinfo{person}{Changshuo Zhang}, \bibinfo{person}{Zihan Lin}, \bibinfo{person}{Shukai Liu}, \bibinfo{person}{Yongqi Liu}, {and} \bibinfo{person}{Han Li}.} \bibinfo{year}{2025}\natexlab{a}.
\newblock \showarticletitle{Comment Staytime Prediction with LLM-enhanced Comment Understanding}.
\newblock \bibinfo{journal}{\emph{arXiv preprint arXiv:2504.01602}} (\bibinfo{year}{2025}).
\newblock


\bibitem[Zhang et~al\mbox{.}(2024d)]%
        {zhang2024qagcf}
\bibfield{author}{\bibinfo{person}{Changshuo Zhang}, \bibinfo{person}{Teng Shi}, \bibinfo{person}{Xiao Zhang}, \bibinfo{person}{Yanping Zheng}, \bibinfo{person}{Ruobing Xie}, \bibinfo{person}{Qi Liu}, \bibinfo{person}{Jun Xu}, {and} \bibinfo{person}{Ji-Rong Wen}.} \bibinfo{year}{2024}\natexlab{d}.
\newblock \showarticletitle{QAGCF: Graph Collaborative Filtering for Q\&A Recommendation}.
\newblock \bibinfo{journal}{\emph{arXiv preprint arXiv:2406.04828}} (\bibinfo{year}{2024}).
\newblock


\bibitem[Zhang et~al\mbox{.}(2025b)]%
        {zhang2025test}
\bibfield{author}{\bibinfo{person}{Changshuo Zhang}, \bibinfo{person}{Xiao Zhang}, \bibinfo{person}{Teng Shi}, \bibinfo{person}{Jun Xu}, {and} \bibinfo{person}{Ji-Rong Wen}.} \bibinfo{year}{2025}\natexlab{b}.
\newblock \showarticletitle{Test-Time Alignment for Tracking User Interest Shifts in Sequential Recommendation}.
\newblock \bibinfo{journal}{\emph{arXiv preprint arXiv:2504.01489}} (\bibinfo{year}{2025}).
\newblock


\bibitem[Zhang et~al\mbox{.}(2024b)]%
        {zhang2024saqrec}
\bibfield{author}{\bibinfo{person}{Kepu Zhang}, \bibinfo{person}{Teng Shi}, \bibinfo{person}{Sunhao Dai}, \bibinfo{person}{Xiao Zhang}, \bibinfo{person}{Yinfeng Li}, \bibinfo{person}{Jing Lu}, \bibinfo{person}{Xiaoxue Zang}, \bibinfo{person}{Yang Song}, {and} \bibinfo{person}{Jun Xu}.} \bibinfo{year}{2024}\natexlab{b}.
\newblock \showarticletitle{SAQRec: Aligning Recommender Systems to User Satisfaction via Questionnaire Feedback}. In \bibinfo{booktitle}{\emph{Proceedings of the 33rd ACM International Conference on Information and Knowledge Management}}. \bibinfo{pages}{3165--3175}.
\newblock


\bibitem[Zhang et~al\mbox{.}(2022)]%
        {zhang2022counteracting}
\bibfield{author}{\bibinfo{person}{Xiao Zhang}, \bibinfo{person}{Sunhao Dai}, \bibinfo{person}{Jun Xu}, \bibinfo{person}{Zhenhua Dong}, \bibinfo{person}{Quanyu Dai}, {and} \bibinfo{person}{Ji-Rong Wen}.} \bibinfo{year}{2022}\natexlab{}.
\newblock \showarticletitle{Counteracting user attention bias in music streaming recommendation via reward modification}. In \bibinfo{booktitle}{\emph{Proceedings of the 28th ACM SIGKDD Conference on Knowledge Discovery and Data Mining}}. \bibinfo{pages}{2504--2514}.
\newblock


\bibitem[Zhang et~al\mbox{.}(2024c)]%
        {zhang2024model}
\bibfield{author}{\bibinfo{person}{Xiao Zhang}, \bibinfo{person}{Teng Shi}, \bibinfo{person}{Jun Xu}, \bibinfo{person}{Zhenhua Dong}, {and} \bibinfo{person}{Ji-Rong Wen}.} \bibinfo{year}{2024}\natexlab{c}.
\newblock \showarticletitle{Model-Agnostic Causal Embedding Learning for Counterfactually Group-Fair Recommendation}.
\newblock \bibinfo{journal}{\emph{IEEE Transactions on Knowledge and Data Engineering}} (\bibinfo{year}{2024}).
\newblock


\bibitem[Zhang et~al\mbox{.}(2020)]%
        {zhang2020learning}
\bibfield{author}{\bibinfo{person}{Yinan Zhang}, \bibinfo{person}{Yong Liu}, \bibinfo{person}{Peng Han}, \bibinfo{person}{Chunyan Miao}, \bibinfo{person}{Lizhen Cui}, \bibinfo{person}{Baoli Li}, {and} \bibinfo{person}{Haihong Tang}.} \bibinfo{year}{2020}\natexlab{}.
\newblock \showarticletitle{Learning personalized itemset mapping for cross-domain recommendation}.
\newblock  (\bibinfo{year}{2020}).
\newblock


\bibitem[Zhao et~al\mbox{.}(2024)]%
        {zhao2024counteracting}
\bibfield{author}{\bibinfo{person}{Haiyuan Zhao}, \bibinfo{person}{Guohao Cai}, \bibinfo{person}{Jieming Zhu}, \bibinfo{person}{Zhenhua Dong}, \bibinfo{person}{Jun Xu}, {and} \bibinfo{person}{Ji-Rong Wen}.} \bibinfo{year}{2024}\natexlab{}.
\newblock \showarticletitle{Counteracting Duration Bias in Video Recommendation via Counterfactual Watch Time}.
\newblock \bibinfo{journal}{\emph{arXiv preprint arXiv:2406.07932}} (\bibinfo{year}{2024}).
\newblock


\bibitem[Zheng et~al\mbox{.}(2023)]%
        {zheng2023reciprocal}
\bibfield{author}{\bibinfo{person}{Bowen Zheng}, \bibinfo{person}{Yupeng Hou}, \bibinfo{person}{Wayne~Xin Zhao}, \bibinfo{person}{Yang Song}, {and} \bibinfo{person}{Hengshu Zhu}.} \bibinfo{year}{2023}\natexlab{}.
\newblock \showarticletitle{Reciprocal sequential recommendation}. In \bibinfo{booktitle}{\emph{Proceedings of the 17th ACM Conference on Recommender Systems}}. \bibinfo{pages}{89--100}.
\newblock


\end{thebibliography}
